\documentclass[aps,prx,twocolumn,nofootinbib,10pt,superscriptaddress,longbibliography]{revtex4-2}
\usepackage{ulem}

\usepackage{amsmath}
\usepackage{amsfonts}
\usepackage{graphicx}
\usepackage{xcolor}
\usepackage{dsfont}
\usepackage[german, english]{babel}
\usepackage{amssymb}
\usepackage{ifthen}
\usepackage{soul}
\usepackage{color}

\usepackage{bm}

\usepackage{adjustbox}

\DeclareMathOperator{\Tr}{Tr}

\def\bkfa{{Ba$_{1-x}$K$_x$Fe$_2$As$_2$}}
\def\MFs{{\mathbf{MF}}}
\def\MF{{\mathbf{MF^s}}}
\def\Ps{{\mathbf{P}}}

\def\od{{00}}
\def\oMFs{{\emptyset}}

\newcommand{\appref}[1]{Appendix~\hyperref[#1]{\ref*{#1}}}
\newcommand{\figref}[1]{Fig. \ref{#1}}
\renewcommand{\eqref}[1]{Eq. (\ref{#1})}
\newcommand{\ii}{\mathrm{i}}

\RequirePackage[colorlinks=true,allcolors=blue,
unicode=true,hypertexnames=false]{hyperref}
\hypersetup{pdfstartview={XYZ null null 1.25}}

\graphicspath{{images/}}

\begin{document}

\title{Microscopic theory of electron quadrupling condensates}

\author{Albert Samoilenka}
\email{albert.samoilenka@gmail.com}
\affiliation{Department of Physics, KTH-Royal Institute of Technology, SE-10691 Stockholm, Sweden}
\affiliation{Department of Physics, Stockholm University, SE-10691 Stockholm, Sweden}

\author{Egor Babaev}
\affiliation{Department of Physics, KTH-Royal Institute of Technology, SE-10691 Stockholm, Sweden}
\affiliation{Wallenberg Initiative Materials Science for Sustainability, Department of Physics, KTH-Royal Institute of Technology, SE-10691 Stockholm, Sweden}

\begin{abstract}
Electron pairing at low temperatures leads to superconductivity. A fundamental question is whether more complex states—characterized by order in four-electron composite objects, termed electron quadrupling or composite order—can exist in materials, and if so, under what conditions they emerge and what properties they exhibit.
These states lie beyond the scope of Bardeen-Cooper-Schrieffer theory, and a microscopic description of them remained elusive.
In the first part of the paper, we provide a general microscopic framework to describe these and the other four-fermion composite states.
In the second part of the paper, we derive and solve a specific fermionic model in two and three dimensions that hosts time-reversal symmetry-breaking electron quadrupling order.
The fermionic microscopic theory is used to estimate the specific heat and electron density of states.
\end{abstract}

\maketitle

\section{Introduction}

At a microscopic level, electron pairing and the resulting superconducting state were described by the Bardeen-Cooper-Schrieffer (BCS) theory \cite{cooper1956bound,Bardeen1957a,Bardeen1957b}.
The superconducting order parameter is a composite of two electronic operators, that we denote by $f$.
Namely it is given by some linear combination of $\langle f_x f_y \rangle$, where, in general the indices $x, y$ can denote space coordinate, band, spin, etc.
The frequently used example is $\langle f_{\uparrow \alpha} f_{\downarrow \alpha} \rangle$, where $\alpha$ is a band index and space coordinate is omitted for brevity.
Recent experiments \cite{grinenko_state_2021,shipulin_calorimetric_2023,bartl2025evidence,halcrow_probing_2024} on \bkfa \ suggest the breaking of Time-Reversal Symmetry (TRS) above the superconducting transition, which was interpreted as the formation of a more complicated state of matter than pairing.
Namely, in this case, the order arises only for fields composed of four electronic operators given by a linear combination of $\langle f_{\uparrow \alpha} f_{\downarrow \alpha} f_{\downarrow \beta}^\dagger f_{\uparrow \beta}^\dagger \rangle$.
This naturally raises the question of how to understand the microscopic physics of such states in a general framework.

We emphasize that in general, the situation is more complex: if one considers objects composed of four electronic operators, a wide range of orders can be realized.
In this work, we start with a general model that allows different four-fermion order parameters, breaking various symmetries, such as linear combinations of $\langle f_x f_y f_z f_w \rangle$ or $\langle f_x f_y f_z^\dagger f_w^\dagger \rangle$.
That includes, e.g., composite orders with nontrivial spin or nonlocal in space part; hence, in the first part of the paper, when discussing the generic formalism, we keep the indices ${}_{x,y,z,w}$ in the most general form, that goes beyond the typical band indices.
In the second part of the paper, we specialize in the especially interesting case where TRS is spontaneously broken, as a result of correlations between pairs of electron pairs with the order parameter given by the linear superposition of $\langle f_{\uparrow \alpha} f_{\downarrow \alpha} f_{\downarrow \beta}^\dagger f_{\uparrow \beta}^\dagger \rangle$ corresponding to the $Z_2$ symmetry.
Such a state involves order in the phase differences between non-condensed electron pairs in different bands (for a phenomenological description, see \cite{bojesen_time_2013,bojesen_phase_2014,maccari_effects_2022}).
This state possesses a number of unprecedented transport and thermoelectric properties.
In the classical-field-theory-based Monte-Carlo calculations  \cite{bojesen_time_2013,bojesen_phase_2014,maccari_effects_2022}), it forms at intermediate temperatures, where at lower temperatures the system is a superconductor that breaks TRS.
As the system is heated, firstly a phase transition takes place at which the superconducting order associated with bilinear order parameters $\langle f_{\uparrow \alpha} f_{\downarrow \alpha} \rangle$ is lost.
However, in that scenario, the system retains order in higher-electronic correlators that results in a state that breaks TRS and is described by an order parameter given by the linear combination of $\langle f_{\uparrow \alpha} f_{\downarrow \alpha} f_{\downarrow \beta}^\dagger f_{\uparrow \beta}^\dagger \rangle$.

This type of order is impossible to describe within the framework of BCS theory, since the relevant quantity is fourth order in fermionic operators (hence termed quadrupling), and the BCS framework for treating the interaction part of the standard microscopic Hamiltonian would not be possible.

The previous theoretical demonstrations of such states were based on phenomenological classical field theory -- Ginzburg-Landau or London models.
Namely, the following  works considered the case where at low or zero temperature the system is a superconductor \cite{babaev2002phase,babaev_superconductor_2004,bojesen_time_2013,bojesen_phase_2014,agterberg_dislocations_2008,how_broken_2024,maccari_effects_2022,
kuklov_deconfined_2006,kuklov_deconfined_2008,
radzihovsky_quantum_2009,berg_charge-4e_2009,herland_phase_2010} (for a brief review see \cite{svistunov_superfluid_2015}).
Counterparts of this state in purely bosonic models were studied in \cite{kuklov_counterflow_2003,kuklov_commensurate_2004,zheng_counterflow_2025,soyler_sign-alternating_2009,sellin_superfluid_2018}.
Within the BCS theory for $U(1)\times Z_2$ (or some other, like $U(1)\times U(1)$) superconductor, generically, the highest critical temperature is associated with the superconducting transition or transition where the system becomes superconducting, simultaneously breaking additional symmetries.
Namely, the other symmetries, such as time-reversal $Z_2$, can only be broken at the same or lower temperature $T_{BTRS} \leq T_{SC}$ (here and below BTRS denotes broken time-reversal symmetry and SC - superconductivity), and hence no composite order is possible within BCS theory framework.

The previous works, that proposed existence of composite order in superconducting systems relied on a two-step approach.
First step was deriving a BCS theory at low temperature and Ginzburg-Landau or London field theory that follows from that or introducing a phenomenological classical field theory of the superconducting state.
Then, the second step introduced fluctuation corrections to such a mean-field theory.
Within this approach, it was shown that under certain restrictive conditions, the sequence of the phase transitions as a function of temperature can change.
For example, a so-called $s+is$, $s+id$ or $p+ip$ superconductor which breaks $U(1)\times Z_2$ symmetry at low temperature, can have $T_{SC} < T_{BTRS}$ \cite{bojesen_time_2013,bojesen_phase_2014,maccari_effects_2022}.
In that case, the intermediate electron quadrupling phase is described by a product of two classical fields, hence, effectively by a four-electron order parameter, given by the linear combination of $\langle b_x b_y^\dagger \rangle$, where the bosonic field $b$ corresponds to a superposition of pairs of electrons in different bands.
In that construction, one can make conclusions only on classical-field-theory aspects of the electron quadrupling state.
Similarly, a system with $U(1)\times U(1)$ symmetry can have an electron quadrupling state that spontaneously breaks $U(1)$ corresponding to a phase difference between the components.
In that case, the corresponding massless mode leads to dissipationless counterflow of the components \cite{babaev2002phase,babaev_superconductor_2004,kuklov_deconfined_2006,herland_phase_2010}.

Fermionic properties of a physical state, such as the quasiparticle spectrum, are crucial for understanding many of its physical responses, like transport, thermal, ultrasound attentuation, Nuclear Magnetic Resonance and thermoelectric behavior of the material.
The BCS theory successfully provided a microscopic framework to describe the fermionic properties of superconductors.
However, no microscopic fermionic theory has been reported for states involving electron quadrupling.
This paper presents the development of such a theory.

\section{Generic model of pair and quadruple condensates of fermions}\label{sec_generic_model}

We start from a generic model of interacting fermions (represented by Grassmann numbers $f$), which is given by the action:
\begin{equation}\label{S_f}
S_f = K_{x y} f_x^\dagger f_y + (v_{x y z w} + W_{x y z w}) f_x^\dagger f_y^\dagger f_z f_w
\end{equation}
where repeated indices are summed/integrated over and include component (spin and band) index, imaginary time, and real space coordinate (or frequency and momentum), such that $x = (\sigma, \alpha, \tau, \textbf{r})$ ($\textbf{r}$ is spatial coordinate that can be continuous or discrete).
Interaction is separated into two parts: $v$ is a pairing interaction that leads to superconductivity, and the remaining part is $W$, which, for example, can correspond to some repulsion.
Namely, we choose them such that when diagonalized in terms of pairs of fermions $v$ ($W$) will have negative (positive) eigenvalues.
We consider a generic model that can have the standard form of interaction between fermions, such as Coulomb, electron-phonon or other type of interaction. 
Recall that the standard BCS mean-field approximation does not permit quadrupling condensate.
Neither can BCS approach be straightforwardly generalized to one supporting quadrupling condensates in the model \eqref{S_f}, since this mean-field should be quartic in fermionic fields $f$'s.

We first perform a Hubbard–Stratonovich transformation, which effectively lowers the number of fermionic fields in the interaction term.
We introduce auxiliary bosonic fields $b$, represented by usual complex-valued numbers.
The corresponding boson is related to a pair of fermions.
With this we obtain the following model:
\begin{equation}\label{S_fb}
\begin{gathered}
S_{fb} = K_{x y} f_x^\dagger f_y + E_{x y} b_x^* b_y \\
+ V_{x y z} b_x^* f_y f_z + V_{x z y}^* b_x f_y^\dagger f_z^\dagger + W_{x y z w} f_x^\dagger f_y^\dagger f_z f_w
\end{gathered}
\end{equation}
Alternatively, this model can also be viewed as a starting point, and hence we will keep the general form of $E$ and $V$.
Note that component indices for bosons are, in general, different from fermionic components.
Here we will consider the Hermitian model \eqref{S_fb}, however, if Hubbard–Stratonovich transformation is used for repulsive interaction terms, the resulting auxiliary bosonic fields would not nucleate superconductivity.
Moreover, it would produce non-Hermitian terms: in that case, $V^*$ won't be a complex conjugate of $V$.

We introduce generic mean-fields $C,\ P,\ d,\ U,\ \Delta$ and separate action into $S_0$ and correction $\delta S$.
Such that $S = S_0 + \delta S$ and $S_0 = S_0^f + S_0^b$, with:
\begin{equation}
\begin{gathered}
S_0^b = (E_{x y} + C_{x y}) b_x^* b_y + P_{x y} b_x^* b_y^* + P_{x y}^* b_x b_y + d_x b_x^* + d_x^* b_x \\
S_0^f = (K_{x y} + U_{x y}) f_x^\dagger f_y + \Delta_{x y} f_x^\dagger f_y^\dagger + \Delta_{y x}^* f_x f_y
\end{gathered}
\end{equation}
Here $U$ mean-field renormalizes the standard Green's function for electrons and can, for example, describe a transition to a ferromagnetic phase when it breaks the corresponding symmetry.

The mean fields $\Delta$ introduce anomalous Green's functions and, when non-zero, signal a transition into a superconducting state, as they break the global $U(1)$ symmetry.
The conventional BCS approach involves a non-zero $\Delta$ (and sometimes also a non-zero $U$) while neglecting other mean fields.
The mean-field denoted by $d$ is also associated with superconductivity; however, it can be removed if another term is introduced into the expansion, see \appref{app_SC_b_f} for details.

By $P$ and $C$ we denote mean-fields that correspond to various, so-called, quadrupling (in fermions), sometimes also termed composite or vestigial orders.
That includes the composite order stabilized by fluctuations: i.e., the electron quadrupling condensates that arise from fluctuation-driven partial restoration of symmetry of a superconducting condensate.
Namely, non-zero values of $P$ correspond to $4e$ superconductivity, described by $\langle f_x f_y f_z f_w \rangle$, since they signal condensation of pairs of bosonic fields $b$, where each bosonic particle relates to a pair of electrons.
$C$ mean-fields lead to renormalization of the bosonic propagators, and hence, when they break corresponding symmetry, $C$ fields can lead to fermion quadrupling condensates of $\langle f_x f_y f_z^\dagger f_w^\dagger \rangle$ type.
In the next section of this work, we will focus on the latter quadrupling states.

So, overall, the model presented here can account for multiple broken symmetries.
Moreover, this approach also applies to the description of condensates beyond four-electron order: e.g. $6e$, $8e$, etc.
Note that composite orders, characterized by more than $2$ fermionic operators, can be formed in different ways.
For example, quadrupling order given by $\langle f_x f_y f_z^\dagger f_w^\dagger \rangle$ can be induced by preformed non-condensed electron pairs $f_x f_y$, which break some higher symmetry due to correlation between components (as we will see below).
Alternatively, one can imagine a different situation: a state with preformed excitons $f_x f_y^\dagger$ which then condense in pairs.
Motivated by recent experiments we are interested in a quadrupling state that can appears above superconducting phase transition, and hence we focus on appropriate Hubbard–Stratonovich transformation.
For higher-order condensates the number of possibilities grows.
One option for condensates of $2 k$ order in fermions is to separate the interaction $(f^\dagger)^k f^k$ (dropping indcies for brevity) into $b (f^\dagger)^k + b^\dagger f^k$ with some auxilary field $b$.
Then one can study states which are defined by order parameter quadratic in $b$'s.

In order to properly define expansion in terms of $\delta S$ we introduce auxiliary variable $\xi$ such that $S(\xi=0) = S_0$ is quadratic in $f$ and $b$, while $S(\xi=1) = S_{fb}$ is the initial physical model \eqref{S_fb}.
Then $\delta S$ is given by:
\begin{equation}
\begin{gathered}
\delta S = \xi V_{x y z} b_x^* f_y f_z + \xi V_{x z y}^* b_x f_y^\dagger f_z^\dagger + \xi W_{x y z w} f_x^\dagger f_y^\dagger f_z f_w\\
- C_{x y} b_x^* b_y - P_{x y} b_x^* b_y^* - P_{x y}^* b_x b_y - d_x b_x^* - d_x^* b_x \\
- U_{x y} f_x^\dagger f_y - \Delta_{x y} f_x^\dagger f_y^\dagger - \Delta_{y x}^* f_x f_y
\end{gathered}
\end{equation}
where all the introduced mean-fields (counterterms) are expanded in powers of $\xi$ up to desired orders.
So, for example, counterterms corresponding to fermion quadrupling orders are given by $C_{x y} = \sum_{i=1}^N \xi^i C_{x y}^i$.
We will call mean-fields in $\delta S$ counterterms, where it is needed to distinguish them from the mean-fields in the $S_0$.
Note that there is actually much freedom in defining such a model.
Namely, the model can account for various phases as well as it can be tuned to better capture some particular phase: (i) prefactors for terms with $V$ and $W$ can be different functions of $\xi$ in general, (ii) one can set some of the mean-fields to desired values (even to zero) and not solve for them self-consistently, when, for example, given mean-field doesn't contribute to the phase transition of interest (iii) if it is related to other phases, higher order interaction terms can be added, provided that they are zero for $\xi = 0$ and $1$.
Overall, this approach of adding and subtracting terms in other contexts is known as shifted-action \cite{shifted_action}, while more generic changes constitute the homotopic action formalism \cite{homotopic_action}.
Both of them are used in various problems to controllably evaluate perturbative expansion in $\xi$ using Diagrammatic Monte-Carlo approach.

Note that one can, in principle, integrate out the auxiliary bosonic fields to go back to the fermionic model.
This would make the prefactors of terms quadratic and quartic in fermionic operators non-trivial functions of $\xi$.
Mainly, it would correspond to interaction terms with some specific prefactors, which are more complicated functions of $\xi$.

We will denote all the introduced mean-fields:
\begin{equation}\label{MFs}
\MFs = (C_{x y}^1, C_{x y}^2, ..., C_{x y}^N, P_{x y}^1, ..., d_x^1, ...)
\end{equation}
The thermal average of any observable $A$ can be computed as a path integral:
\begin{equation}\label{aver_A}
\begin{gathered}
\langle A \rangle = \frac{\int \mathcal{D}[f, f^\dagger, b, b^*] A e^{-S}}{Z} \\ 
Z = \int \mathcal{D}[f, f^\dagger, b, b^*] e^{-S}
\end{gathered}
\end{equation}
where $Z$ is the partition function and one expands in powers of $\xi$ up to some order $N$ to obtain connected Feynman diagrams, which can be evaluated since $S_0$ is quadratic in bosonic and fermionic operators.
Note that since bosonic fields are regular complex-valued fields, it is assumed that mean-fields are chosen such that corresponding Gaussian integrals converge.

We will use the following notation for propagators:
\begin{equation}\label{P_defs}
\begin{gathered}
\langle f_x f_y \rangle_0 = F_{x y} =
\adjincludegraphics[valign=c,width=0.25\linewidth]
{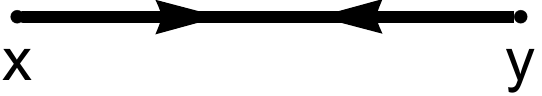}
\\
-\langle f_x^\dagger f_y^\dagger \rangle_0 = F_{x y}^* =
\adjincludegraphics[valign=c,width=0.25\linewidth]
{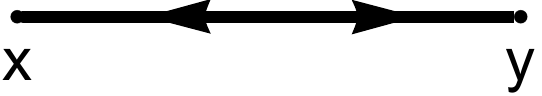}
\\
-\langle f_x f_y^\dagger \rangle_0 = G_{x y} = \adjincludegraphics[valign=c,width=0.25\linewidth]
{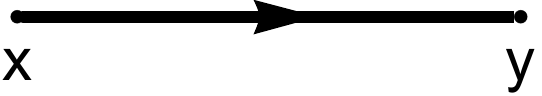}
\\
-\langle f_x f_y^\dagger \rangle_\oMFs = g_{x y} =
\adjincludegraphics[valign=c,width=0.25\linewidth]
{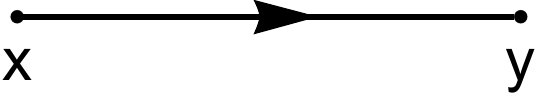}
\\
\langle b_x b_y \rangle_\od = T_{x y} =
\adjincludegraphics[valign=c,width=0.25\linewidth]
{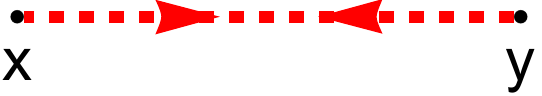}
\\
\langle b_x^* b_y^* \rangle_\od = T_{x y}^* =
\adjincludegraphics[valign=c,width=0.25\linewidth]
{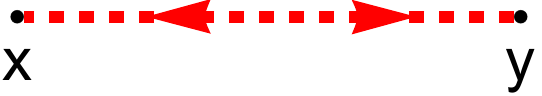}
\\
\langle b_x b_y^* \rangle_\od = R_{x y} =
\adjincludegraphics[valign=c,width=0.25\linewidth]
{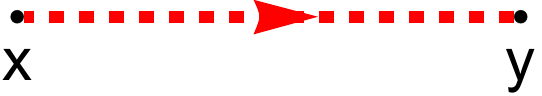}
\\
\langle b_x b_y^* \rangle_\oMFs = r_{x y} =
\adjincludegraphics[valign=c,width=0.25\linewidth]
{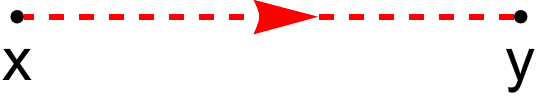}
\\
\langle b_x \rangle_0 = \Psi_x = \raisebox{0.6ex}{\adjincludegraphics[valign=c,width=0.05\linewidth]
{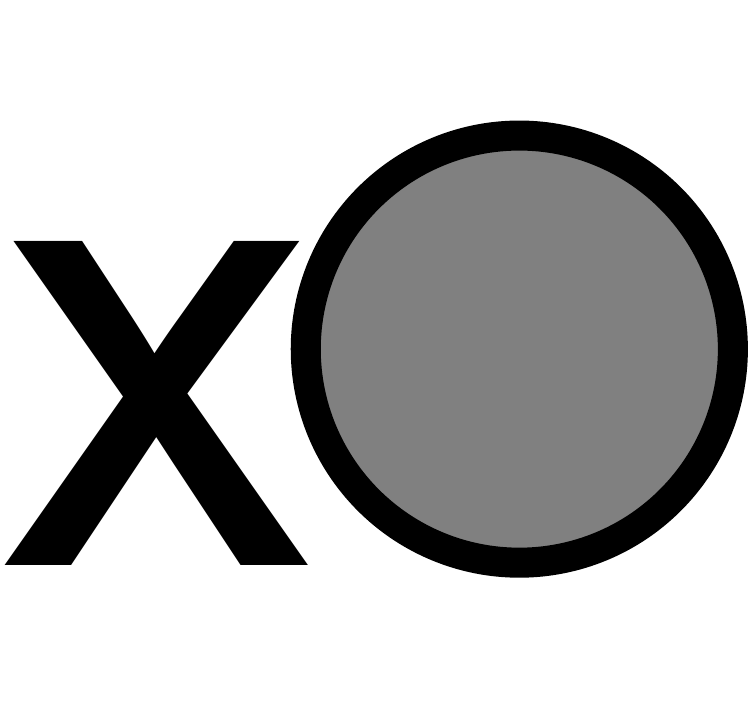}}
\end{gathered}
\end{equation}
where $\langle ... \rangle_0$ is thermal average as in \eqref{aver_A} but with action $S_0$, while $\langle ... \rangle_\od = \left. \langle ... \rangle_0 \right|_{d \rightarrow 0}$ and $\langle ... \rangle_\oMFs = \left. \langle ... \rangle_0 \right|_{\MFs \rightarrow 0}$.
For the interaction vertices, we will adopt the convention that subscripts $x, y, z, ...$ of the vertex are enumerated clockwise in the diagram.

The grand canonical potential is given by:
\begin{equation}\label{Omega}
\Omega = - T \ln Z
\end{equation}
where $T$ is the temperature.
All the mean-fields $\MFs$ are found as extremums of the $\Omega$, which is often referred to as the variational perturbation theory \cite{feynman1986effective,kleinert1998systematic},
meaning that they satisfy the self-consistency equations:
\begin{equation}\label{MFs_eqs}
\frac{\delta \Omega}{\delta \MFs} = 0
\end{equation}

This model provides a setup for calculations in generalized mean-field approximation as well as numerical evaluation of higher-order diagrams using diagrammatic Monte Carlo.
In the latter case, it is related to the variational approach used in \cite{chen_combined_2019}.

\subsection{Alternative form of self-consistency equations and grand canonical potential}

The downside of the model at this point is that \eqref{MFs_eqs} are rather complicated when evaluated.
So let us derive simpler expressions for self-consistency equations \eqref{MFs_eqs} and grand canonical potential \eqref{Omega}.
To that end, consider one of the mean-fields, denoted by $\alpha$, and the corresponding bosonic or fermionic fields prefactor, denoted $A$.
Namely, $S_0 = ... + \alpha A$ with $\alpha = \sum_{i = 1}^N \alpha_i$ and $\delta S = ... - \alpha' A$ with $\alpha' = \sum_{i = 1}^N \xi^i \alpha'_i$, where for the moment we will treat $\alpha_i$ and counterterms $\alpha'_i$ as different variables and set $\alpha_i = \alpha'_i$ at the end of the calculation.
Then since
\begin{equation}\label{deriv_to_aver}
\frac{1}{T} \frac{\delta\Omega}{\delta \alpha_i} = \langle A \rangle,\ \ \ \frac{1}{T} \frac{\delta\Omega}{\delta \alpha'_i} = - \xi^i \langle A \rangle
\end{equation}
self-consistency equations \eqref{MFs_eqs} corresponding to $\alpha_i$ for $i = 1, ... N$ are given by:
\begin{equation}\label{MFs_eqs_deriv}
\frac{1}{T} \frac{\delta\Omega}{\delta \alpha_i} + \frac{1}{T} \frac{\delta\Omega}{\delta \alpha'_i} = \langle A \rangle - \xi^i \langle A \rangle = 0
\end{equation}
Next, all expressions are to be expanded in powers of $\xi$ up to order $N$, such that
\begin{equation}
\langle A \rangle \simeq \langle A \rangle_{\overline{N}} \equiv \langle A \rangle_0 + \xi \langle A \rangle_1 + ...\  \xi^N \langle A \rangle_N
\end{equation}
Since $\xi^i \langle A \rangle \simeq \xi^i \langle A \rangle_{\overline{N-i}}$ self-consistency equations \eqref{MFs_eqs_deriv} become:
\begin{equation}\label{MFs_eqs_avers}
\langle A \rangle_i = 0,\ \ \text{for}\ \ i \in [1,N]
\end{equation}
These equations are derived by expanding in powers of $\xi$ and then setting $\xi = 1$.
The last step is needed to get back to the physical model $S_{fb}$ \eqref{S_fb}.

Using \eqref{deriv_to_aver} and expanding $\Omega = \sum_{i = 0}^N \xi^i \Omega_i$, self-consistency equations \eqref{MFs_eqs_avers} become $\frac{\delta \Omega_i}{\delta \alpha} = 0$.
Since $\alpha$ (distinct from $\alpha'$ here) can be present only in propagators $\Ps$, defined in \eqref{Props} and \eqref{P_defs}, we obtain yet another alternative expression for self-consistency equations:
\begin{equation}\label{MFs_eqs_Ps}
\frac{\delta \Omega_i}{\delta \Ps} = 0,\ \ \text{for}\ \ i \in [1,N]
\end{equation}
where
\begin{equation}\label{Props}
\Ps = (R_{x y}, T_{x y}, \Psi_x, G_{x y}, F_{x y})
\end{equation}

Finally, the self-consistency equations and grand canonical potential can be further simplified by making use of skeleton diagrams -- diagrams that cannot be made disconnected by cutting $2$ fermionic or $1$ or $2$ bosonic lines.

Firstly, we can deduce that counterterms, such as $-\xi^k \alpha'_k A$, can cancel corresponding proper self-energy insertions when calculating thermal averages of any observables or partition function in diagrams to all orders.
When expanding in powers of $\xi$, the action in $e^{-S}$, we will obtain all the interaction and counter terms with prefactors $1/n!$ for some power $n$ of the corresponding term.
Next, the calculation of the thermal average using thermal Wick's theorem will produce all the possible thermal averages of pairs and singles of fields.
Note that accounting for all the permutations of terms, each diagram will be encountered exactly $n!$ times for each term to the power $n$.
This cancels the $1/n!$ prefactors.
Hence, each counterterm insertion can be factored out together with the same proper self-energy term, and hence can cancel it everywhere.

This is also true for the simplest diagram with just one counterterm: for a counterterm of order $k$ that would be the self-consistency equation $\langle A \rangle_k = 0$ \eqref{MFs_eqs_avers}.
Hence, if self-consistency equations are satisfied, then corresponding counterterms will cancel all the proper self-energy insertions, meaning that all diagrams become skeleton diagrams.
It means that we can rewrite the self-consistency equations \eqref{MFs_eqs_Ps} (or \eqref{MFs_eqs_avers}) in the form $- \alpha_i + (\text{i-th order skeleton diagrams}) = 0$, where in skeleton diagrams lower order counterterms were canceled by corresponding proper self-energies.

Similarly, cancellation happens in grand canonical potential $\Omega$.
In a skeleton version of the grand canonical potential $\Omega^s$, we keep the counterterms only in the lowest order: $- \xi^i T \alpha'_i \langle A \rangle_0$.
So we can replace the initial grand canonical potential \eqref{Omega} with $\Omega^s$, by keeping only skeleton diagrams in the expansion.
Corresponding self-consistency equations are $\frac{\delta \Omega^s_i}{\delta \Ps} = 0$.

Note that after setting $\alpha_i = \alpha'_i$ and $\xi = 1$, the skeleton grand canonical potential $\Omega^s$ depends only on the sum of the mean-fields $\alpha$.
We will denote these sums of mean-fields as $\MF = (C_{x y},\ P_{x y},\ d_x,\ U_{x y},\ \Delta_{x y})$.
Moreover, since all counterterms cancel with proper self-energies, the calculation of any observable also depends only on $\MF$.
Whereas self-consistency equations can be combined to produce equations only for $\MF$:
\begin{equation}\label{MFs_eqs_s}
\frac{\delta (\Omega^s - \Omega_0)}{\delta \Ps} = 0, \ \ \text{or equivalently} \ \ \frac{\delta \Omega^s}{\delta \MF} = 0
\end{equation}
This reduces the number of variables, since now we can keep track of only the sum of the mean-fields given by $\MF$.
So we can study the skeleton version of the model $\Omega^s$ and self-consistency equations \eqref{MFs_eqs_s} instead of \eqref{Omega} and \eqref{MFs_eqs}.
The procedure is related to how Luttinger–Ward and Baym–Kadanoff functionals are constructed in other contexts.

\subsection{The model for quadrupling to normal state transition}

Let us consider one of the simplest versions of the model described above that can be used to describe normal and quadrupling states.
To that end, we discard the irrelevant mean-fields $P, d, \Delta, U$.
For later convenience, we also absorb the simplest diagram with single $W$ into the chemical potential for fermions (this can also be viewed as setting $U^1$ equal to this diagram).
Then expanding in powers of $\xi$ and setting it to one gives the skeleton grand canonical potential:
\begin{equation}\label{Omega_diags}
\begin{gathered}
\frac{\Omega^s}{T} = - \Tr \ln R - \Tr(C R) + 2 \hspace{0.2em}\adjincludegraphics[valign=c,width=0.25\linewidth]{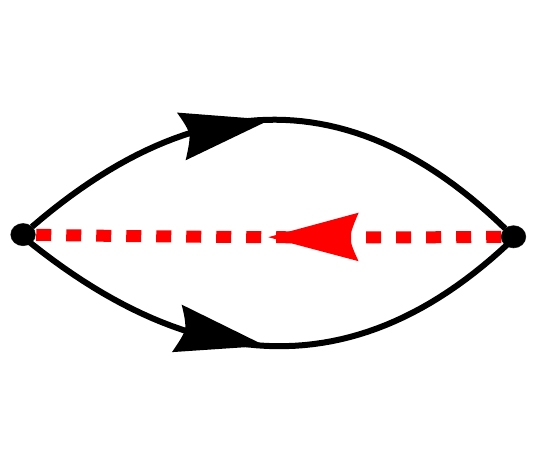}
+ ... \\
+ 8 \hspace{-0.05em}\adjincludegraphics[valign=c,width=0.25\linewidth]{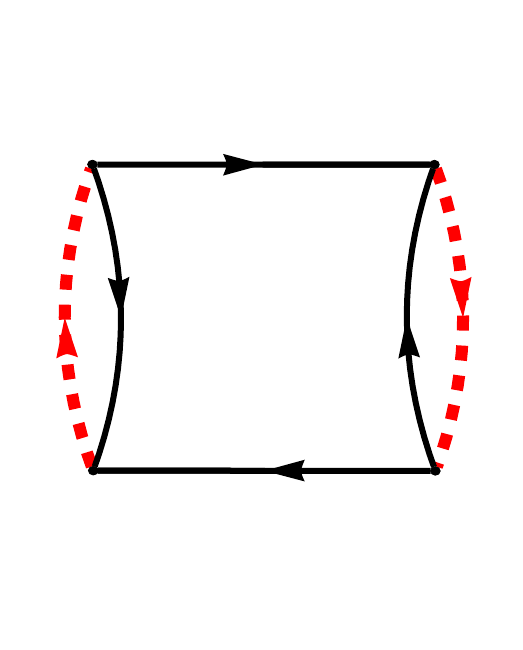}\hspace{-0.2em}
+ ... + \frac{32}{3} \adjincludegraphics[valign=c,width=0.25\linewidth]
{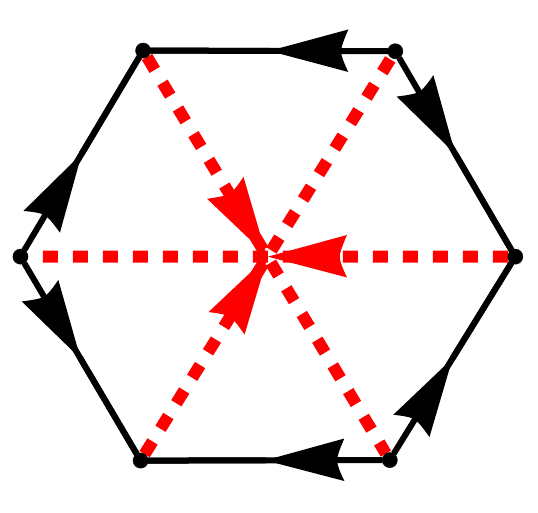} + ... \\
- 128 \hspace{0.1em}\adjincludegraphics[valign=c,width=0.25\linewidth]
{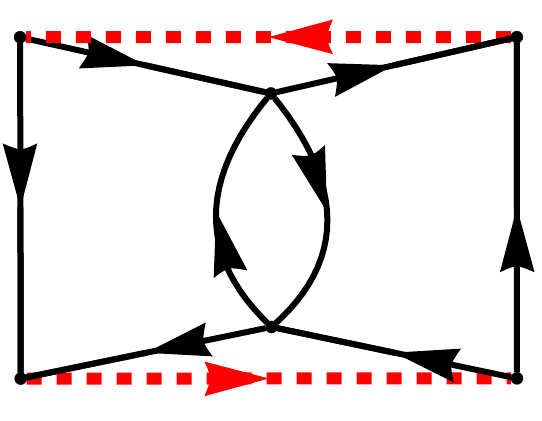} + 64 \adjincludegraphics[valign=c,width=0.25\linewidth]
{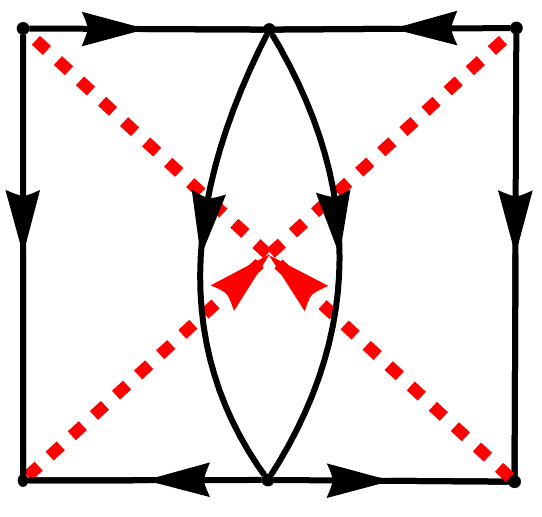} + ...
\end{gathered}
\end{equation}
where we draw only some of the typical diagrams.
The first one is responsible for the usual BCS estimate for the transition into the superconducting state.
Second and third are higher-order in $R$ corrections to it, which, along with many other (not shown here) diagrams, lead to fluctuation corrections (due to the pair propagator) to the transition.
While the last two, as explained below, are some of the diagrams that can lead to the BTRS electron quadrupling state (also called BTRS quartic metal).

\section{Quadrupling transition in a model with density-density interaction}

In this section, we specialize in a concrete
example, motivated by \bkfa.
At low temperature, the material is superconducting with three dominant bands having approximately equal gaps.
We aim at simplifying the model, still retaining certain key features.
To that end, we will consider a continuous model, in two and three dimensions with $3$ symmetric fermionic bands, given by the action:
\begin{equation}
\begin{gathered}
S_f = \int d\tau d\textbf{r} \left[ \sum_{\sigma = \uparrow, \downarrow} \sum_{\alpha = 1}^3 f_{\sigma \alpha}^\dagger \left(\partial_\tau - \frac{\nabla^2}{2 m} - \mu\right) f_{\sigma \alpha} \right. \\
\left. + \sum_{\alpha,\beta = 1}^3 \left( f_{\downarrow \alpha}^\dagger f_{\uparrow \alpha}^\dagger v_{\alpha \beta} f_{\uparrow \beta} f_{\downarrow \beta} + \sum_{\sigma,\sigma' = \uparrow, \downarrow} f_{\sigma \alpha}^\dagger f_{\sigma' \beta}^\dagger W_{\alpha \beta} f_{\sigma' \beta} f_{\sigma \alpha} \right)\right]
\end{gathered}
\end{equation}
where $v$ is the interaction that leads to superconductivity and $W$ is a repulsive density-density interaction with diagonal elements assumed to be absorbed in the definition of $v$, which means we set $W_{\alpha \alpha} = 0$.
Moreover, only the sum of $W$'s will contribute, so we put $W_{\alpha \beta} = W > 0$.

We fix the $v$ interaction to $v_{\alpha \alpha} = -2 \lambda / 3$ and $v_{\alpha \neq \beta} = \lambda/3$ with $\lambda > 0$ such that it is diagonalized by $v_{\alpha \beta} = -\lambda \sum_{a = 1,2} e_{a \alpha}^* e_{a \beta}$ with $e_{1 \alpha} = (1, e^{2 \pi \ii / 3}, e^{- 2 \pi \ii / 3}) / \sqrt{3}$ and $e_{2 \alpha} = e_{1 \alpha}^*$.

Note that for two degenerate eigenvalues $- \lambda$, there is freedom in the choice of corresponding eigenvectors.
Moreover, in general, there is freedom in how auxiliary bosonic fields $b$ are chosen: we can transform $b_x \to B_{x y} b_y$ for some unitary $B$.
Based on these choices TRS transformation is defined differently.
For example, when $V_{x y z}$ ($e_{a \alpha}$ in the current model) is real, then TRS transformation corresponds to complex conjugation of all scalars in the action ($f$ and $b$ fields are not scalars).
Hence, the state has TRS, if $C,R \in \text{Re}$ (ignoring other mean-fields), and TRS is broken if they are complex.
In two component model BTRS corresponds to $C_{1 2} \neq C_{1 2}^*$ (same for $R$), since $C_{1 1}, C_{2 2} \in \text{Re}$ always.
In other words, BTRS is interpreted as a state when the difference between phases of different bosonic components is not $0$ or $\pi$.

\begin{figure*}
\centering
\includegraphics[width=1\linewidth]{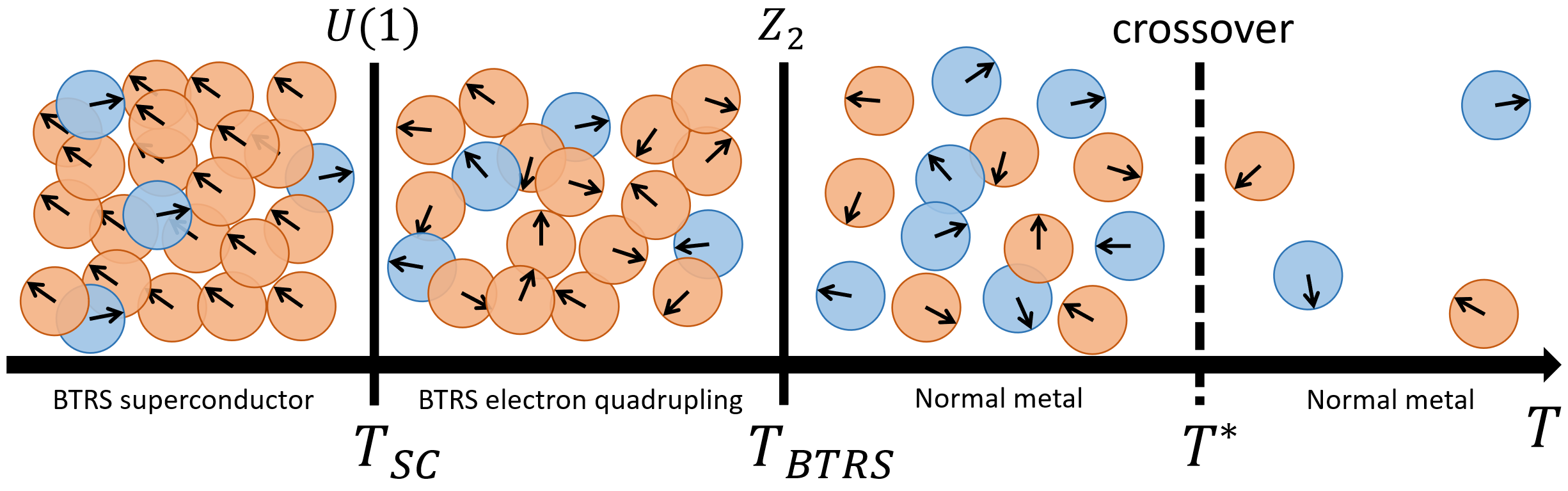}
\caption{
Cartoon of different states in the studied system.
Orange (blue) colored disks denote $s + \ii s$ ($s - \ii s$) Cooper pairs, with arrow schematically indicating corresponding phase $[0, 2 \pi]$.
Going from right (high temperature) to left (low temperature), we depict the following states:
($T^* \approx T_{BCS} < T$) normal metal with very few preformed Cooper pairs with uncorrelated phases and an equal number of $s \pm \ii s$ pairs (here $T_{BCS}$ should be understood as a characteristic temperature scale, not a temperature of a phase transition);
($T_{BTRS} < T < T^*$) resistive metal but with stronger local ordering effects and an increased, but still equal number of preformed $s \pm \ii s$ pairs;
($T_{SC} < T < T_{BTRS}$) BTRS electron quadrupling state where symmetry between the number of $s \pm \ii s$ pairs is spontaneously broken -- meaning that there is a different number of $s + \ii s$ and $s - \ii s$ pairs, while superconducting phases are still uncorrelated on a long-range scale.
This corresponds to onset of non-zero order parameter quartic in electronic fields $\langle f_{\uparrow \alpha} f_{\downarrow \alpha} f_{\downarrow \beta}^\dagger f_{\uparrow \beta}^\dagger \rangle$.
($T < T_{SC}$) BTRS superconductor -- phases become ordered, symmetry between $s \pm \ii s$ pairs remains spontaneously broken.
}
\label{fig_cartoon}
\end{figure*}

In this work, we chose another equivalent representation where $V_{x y z}$ (and hence $e_{a \alpha}$) are of some specific complex-valued form.
In this case TRS transformation is given by complex conjugation of all scalars and the swap $b_1 \leftrightarrow b_2$.
In a sense, this choice produces a diagonal representation of TRS, since the model has TRS if $C_{1 1} = C_{2 2}$, and TRS is broken otherwise.
So overall this interaction produces $2$ symmetric bosonic components where one of them (for example $R_+ \equiv R_{1 1}$) corresponds to $s + \ii s$ state, while the other ($R_- \equiv R_{2 2}$) to $s - \ii s$.
This means that in this representation of a multicomponent superconducting state, when the components are not equal, the TRS is broken.
In other words, TRS is broken when there is a different number of particles (electron Cooper pairs) in the bosonic components related by time-reversal operation.
Note that here bosonic components are different from electronic bands, and each bosonic component is given by a linear superposition of electron pairs from different bands.
Speaking informally, there could be more ``$s + \ii s$ electron pairs" than ``$s - \ii s$ pairs" or vice versa. 
We will see below that this can also happen in a non-superconducting state, without pairs being condensed, see \figref{fig_cartoon} for a scheme of the states present in this model.
In this case, the flow of electron charge is resistive, but TRS is broken, similarly to the state suggested in \cite{grinenko_state_2021,shipulin_calorimetric_2023}.
It is termed composite order or electron quadrupling since as a relevant order parameter, one can use (denoting only component indices) $\langle b_1^* b_1 - b_2^* b_2 \rangle$, which is quadratic in bosonic fields, and each $b$ corresponds to a pair of electrons, hence the composite order parameter is fourth-order in fermions.

Consider the diagrams contributing to the generic expression for $\Omega^s$ \eqref{Omega_diags}.
Each bosonic propagator connecting electronic lines of the bands $\alpha$ and $\beta$ has terms proportional to $\sum_a e_{a \alpha}^* e_{a \beta} R_{a a}$.
Hence, when the bands are the same $\alpha = \beta$, these terms depend only on $R_{1 1} + R_{2 2}$ and cannot produce a transition into the BTRS quadrupling state.
On the other hand, diagrams, like the last two in \eqref{Omega_diags}, will have terms in which $R$ connects also electronic lines of different electronic bands, thus allowing the quadrupling state.
In fact, this type of diagrams will produce non-zero values of $\Lambda_-$, see below in \eqref{Omega_L_-d} and \eqref{Omega_L_3d}, which is needed to make a disbalance between the bosonic components favorable.

We will now focus on considering this model in the relatively weak coupling limit given by:
\begin{equation}\label{BCS_limit}
\mu \gg \omega_f \gg T_{BCS} \gg \omega_b
\end{equation}
where $\mu$ is chemical potential; $\omega_f$ sets the energy cutoff for electrons near Fermi surface by $\left| \frac{k^2}{2 m} - \mu \right| \leq \omega_f$ and can correspond to Debye frequency or some other cutoff scale; $T_{BCS} = \frac{2 e^{\gamma_E} \omega_f}{\pi} e^{- \frac{1}{N_F \lambda}}$ (where $\gamma_E$ is Euler's constant and $N_F$ is density of state defined below) is used as a characteristic temperature scale corresponding to what would be the critical temperature in usual BCS approximation for normal to superconducting states transition.
The quantity $\omega_b = v_F q_m$, where $v_F$ is the Fermi velocity and $q_m$ is the maximal momentum of electronic pairs (bosonic particles).
So the $\omega_b$ sets the cutoff for energies of electronic pairs by $|v_n|, |v_F q| \leq \omega_b$, where $v_n = 2 \pi T n$ is bosonic Matsubara frequency.
The last inequality in \eqref{BCS_limit} corresponds to the so-called static approximation since it sets $v_n \simeq 0$ (it can be replaced by $\omega_f \gg \omega_b$ in the non-static approximation).

Below, we will use the following rescaled parameters:
\begin{equation}\label{resc_params}
\begin{gathered}
f = \frac{\omega_f}{\mu},\ \ h = \frac{T_{BCS}}{\omega_f},\ \ s = \frac{\omega_b}{T_{BCS}}, \\
w = N_F W,\ \ t = \frac{T}{T_{BCS}}
\end{gathered}
\end{equation}
where in $2d$ $N_F = \frac{k_F}{2 \pi v_F}$ and in $3d$ $N_F = \frac{k_F^2}{2 \pi^2 v_F}$ is the density of states at the Fermi level of free electrons per band and spin and the parameters are assumed to be small, while $t \simeq 1$.
To keep the model convergent we assume that $w < \text{min} \left( \frac{1}{2}, \left( 2 \ln \frac{2 e^{\gamma_E}}{\pi h t} \right)^{-1} \right)$.
The first bound ensures that the system doesn't transition to an excitonic state for too large and positive $w$, while the second ensures that the series in $w$ doesn't diverge due to the transition to a superconducting state for negative $w$ at that value.

Then the skeleton grand canonical potential is obtained as described in the previous section by expanding in powers of $\xi$ up to $6$'s order:
\begin{equation}\label{Omega_quad}
\begin{gathered}
\frac{\Omega^s}{V} = T \int \frac{d \textbf{q}}{(2 \pi)^2} \Tr \left[ \ln\widetilde{R}^{-1}(\textbf{q}) - C(\textbf{q}) \widetilde{R}(\textbf{q}) \right] \\
+ \frac{T}{V} \left(\mathcal{D}_1 + \mathcal{D}_2 + \mathcal{D}_3 \right)
\end{gathered}
\end{equation}
where $V$ is the volume of the system, the trace $\Tr$ is over bosonic components $1, 2$, $\widetilde{R}$ is a pair propagator in momentum space obtained from $(R^{-1})_{a b} = \delta_{a b}  r_a^{-1} + C_{a b}$ with $r_a = \lambda$ and $\mathcal{D}_i$ are diagrams $i$'s order in $R$.
In this model, normal state corresponds to parameters when local minima of the energy function \eqref{Omega_quad} are $\widetilde{R}_{1 1} = \widetilde{R}_{2 2} < +\infty$, while electron quadrupling BTRS state is given by finite $\widetilde{R}_{1 1} \neq \widetilde{R}_{2 2}$, which spontaneously breaks the corresponding $Z_2$ symmetry (equivalently it can be written in terms of $C_{a b}$).
Transition to superconducting state happens when local minima disappear and hence $\widetilde{R}_{a a}(0) \to +\infty$.

To solve the model \eqref{Omega_quad} in the \eqref{BCS_limit} limit described above, we expand in powers of momenta $q < q_m$ of the pair propagators $\widetilde{R}$ and neglect them compared to the momenta of electrons (which is of order Fermi momenta $k_F$) in diagrams of higher order.
Next, terms $\mathcal{D}_1$, which are first order in $R_{a b}$ (they are proportional to $\delta_{a b}$) are absorbed into the $C_{a b}$ and hence $R_{a b}$, such that $(\widetilde{R}^{-1})_{a b} = N_F (\epsilon_{a b} + \delta_{a b} \eta q^2)$, where $\epsilon_{a b} = \delta_{a b} \epsilon + C_{a b} / N_F$ and parameters are defined below in the following subsections, and we assume that new $C_{a b}$ are constant in momentum.

In general, $\epsilon_{a b}$ are represented by a $2 \times 2$ Hermitian matrix.
However, as shown in \appref{app_R_offdiag}, in the current model, with equal bosonic components, off-diagonal terms are zero in the minimum of the grand canonical potential.
So in the following (sub)sections we will consider only diagonal $\epsilon_+ \equiv \epsilon_{1 1}$, $\epsilon_- \equiv \epsilon_{2 2}$ (and corresponding $R_\pm$).

\subsection{$2d$ system}

We begin by considering a two-dimensional case.
As is well known in two dimensions, one needs to carefully treat vortex excitations \cite{berezinsky_destruction_1972,kosterlitz_ordering_1973}.
Furthermore, for $s+is$ systems, there are principal differences compared to usual Berezinskii-Kosterlitz-Thouless that were considered at the level of Monte-Carlo classical field theory in \cite{bojesen_time_2013,maccari_prediction_2023}.
Here, our main focus will be on the state that breaks $Z_2$ symmetry above the superconducting phase transition and hence has a genuine long-range order. To address that, we will instead use a simpler approximate approach, based on a modification of mean-field theory.
Overall, we rescale the model and introduce new variables $L_\pm$, which are rescaled bosonic (electron pair) propagators averaged over momenta (or at $\textbf{r} = 0$ equivalently) so that they are proportional to a number of particles in a given bosonic component:
\begin{equation}\label{params}
\begin{gathered}
L_\pm = 4 \pi \eta N_F\int \frac{d \textbf{q}}{(2 \pi)^2} \widetilde{R}_\pm = \ln\left( 1 + \frac{\kappa}{\epsilon_\pm} \right) \\
\eta = \frac{7 \zeta_3 v_F^2}{2 (4 \pi T)^2},\ \ 
\kappa = q_m^2 \eta = \frac{7 \zeta_3 s^2}{2 (4 \pi t)^2}
\end{gathered}
\end{equation}
where $\zeta_i$ is Riemann zeta function.
While expression for $\epsilon$ is:
\begin{equation}\label{eps}
\begin{gathered}
\epsilon = \ln t + \frac{f h s^2 w^2}{4 t} (P_2 + P_3 w + P_4 w^2)
\end{gathered}
\end{equation}
where $P_i$ are numerical coefficients that are functions of $\frac{\omega_f}{T} = \frac{1}{h t}$ and defined in \appref{app_Ps}.

Then the model \eqref{Omega_quad} transforms into:
\begin{equation}\label{Omega_L_-d}
\begin{gathered}
\bar{\Omega} \equiv \Omega^s \frac{4 \pi \eta}{T V} = \sum_{a = \pm} \bar{\Omega}_a^0 + \Lambda_2 (L_+ + L_-)^2 \\
- \Lambda_- (L_+ - L_-)^2 - \Lambda_3 (L_+ + L_-)^3 \\
\text{with} \\
\bar{\Omega}_\pm^0 = \kappa\left( \ln\kappa - 1 \right) + (\epsilon + \kappa) L_\pm 
- \kappa \ln\left( e^{L_\pm} - 1 \right) 
\end{gathered}
\end{equation}
where other parameters are given by:
\begin{equation}
\begin{gathered}
\Lambda_2 = \frac{f h t}{3} - \frac{2 P_+ (\pi f h s w)^2}{7 \zeta_3} \\
\Lambda_3 = \frac{31 \zeta_5 (f h t)^2}{147 \zeta_3^2} \\
\Lambda_- = \frac{2 P_- (\pi f h s w)^2}{7 \zeta_3}
\end{gathered}
\end{equation}
$P_\pm$ parameters are defined in \appref{app_Ps}.
Note, that paremeters $\kappa, \Lambda_i, \Lambda_- > 0$.

\begin{figure}
\centering
\includegraphics[width=1\linewidth]{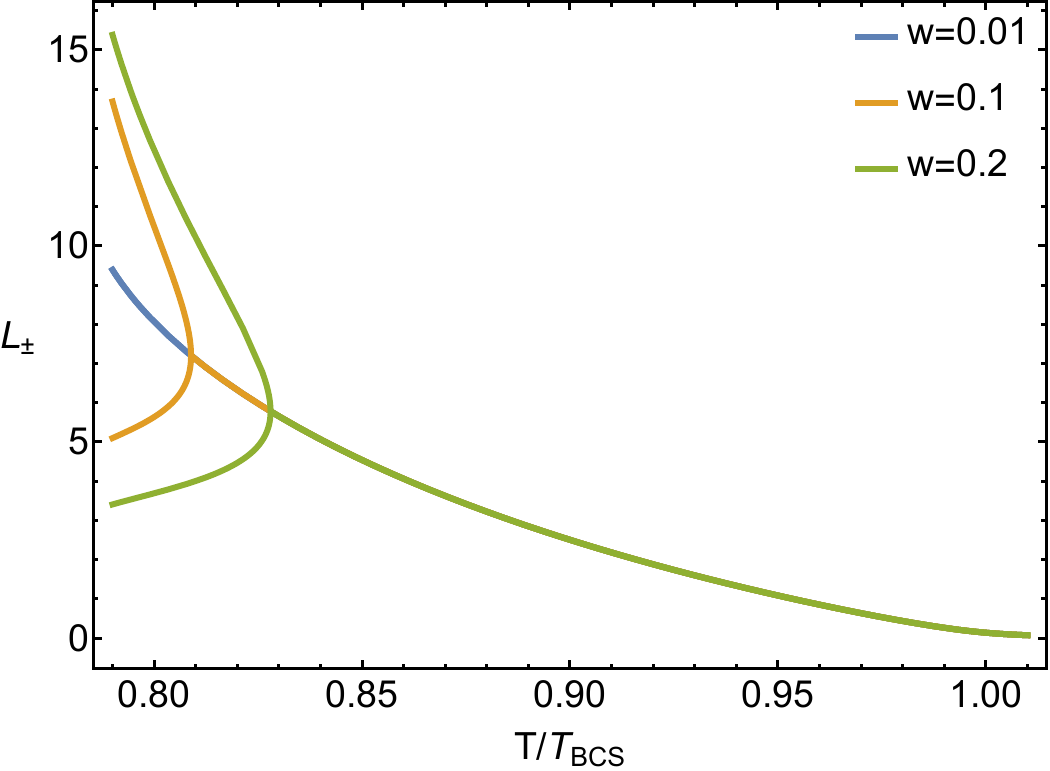}
\caption{
The plot of the rescaled number of electron pairs $L_+,\ L_-$ in the two bosonic components at local minima of the energy function \eqref{Omega_L_-d} as a function of temperature $T$ in $2d$ system for $f = h = s = 0.2$ and different values of the density-density interaction $w$.
Note that as the temperature is decreased at $T^* \simeq T_{BCS}$ $L$'s increase greatly, which signals the pre-formation of Cooper pairs.
A single minimum corresponds to a normal state while splitting into two minima to a BTRS quadrupling state.
For temperatures smaller than shown on the plot, minima disappear and $L$ goes to $+ \infty$, which corresponds to the superconducting state.
When the density-density interaction $w$ is small enough quadrupling state disappears.
So for $w = 0.2$ we have the following phases: superconducting for $t < t_{SC} \simeq 0.783$, quadrupling for $t_{SC} < t < t_{BTRS} \simeq 0.828$ and normal for $t > t_{BTRS}$.
Cooper pairs are preformed for $t < t^* \simeq 1$.
}
\label{fig_Ls_t_2d}
\end{figure}

Let us study the model \eqref{Omega_L_-d}.
$\bar{\Omega}$ always diverges to $- \infty$ for $L_\pm \to + \infty$.
However, for sufficiently large temperatures, there is one or two local minima.
We interpret the state with one minimum as a normal, the state with two minima as quadrupling, and the state with none as a superconducting state, see \figref{fig_Ls_t_2d} for details.
Note that the transition to a superconducting state is given by $L_\pm \to \infty$ or equivalently $\epsilon_\pm \to 0$ for one or both components.

As a side note, if one takes into account only the quadratic $\Lambda_2$ term, then the model always has a minimum, and hence there would be no transition to the superconducting state.

Now, let us consider some properties of these minima.
Firstly, we will consider the case without the BTRS state with a single transition from the normal to the superconducting state.
In normal state single minimum corresponds to $L_+ = L_- = L_0$, which is found from $\frac{d \bar{\Omega}}{d L_0} = 0$ by:
\begin{equation}\label{L0}
L_0 = \frac{\Lambda_2 - \sqrt{\Lambda_2^2 + 3 \Lambda_3 \left( \epsilon - \frac{\kappa}{e^{L_0} - 1} \right)}}{6 \Lambda_3}
\end{equation}
When the temperature is decreased, the expression under the square root decreases, and at some point changes sign to negative.
This temperature $T_{SC} \equiv t_{SC} T_{BCS}$ corresponds to the transition from the normal to the superconducting state.
This transition is suppressed by fluctuations relative to the BCS-mean-field value $T_{BCS}$.
For large $L_0$ and small $w$ it can be estimated from $\epsilon_{SC} \simeq -\frac{\Lambda_2^2}{3 \Lambda_3}$ and $\epsilon \simeq \ln t$, which gives:
\begin{equation}\label{t_SC}
t_{SC} \simeq \exp{\left( -\frac{(7 \zeta_3)^2}{279 \zeta_5} \right)} \simeq 0.783
\end{equation}
Turns out that this estimate stays quite close to numerical solution even for larger $w$ and for the transition from quadrupling to a superconducting state.

Using \eqref{L0} we also obtain the maximal value that $L_0$ can have in the normal state:
\begin{equation}\label{L0max}
L_0^{max} = \left. \frac{\Lambda_2}{6 \Lambda_3} \right|_{t \to t_{SC}}
\end{equation}

\begin{figure}
\centering
\includegraphics[width=1\linewidth]{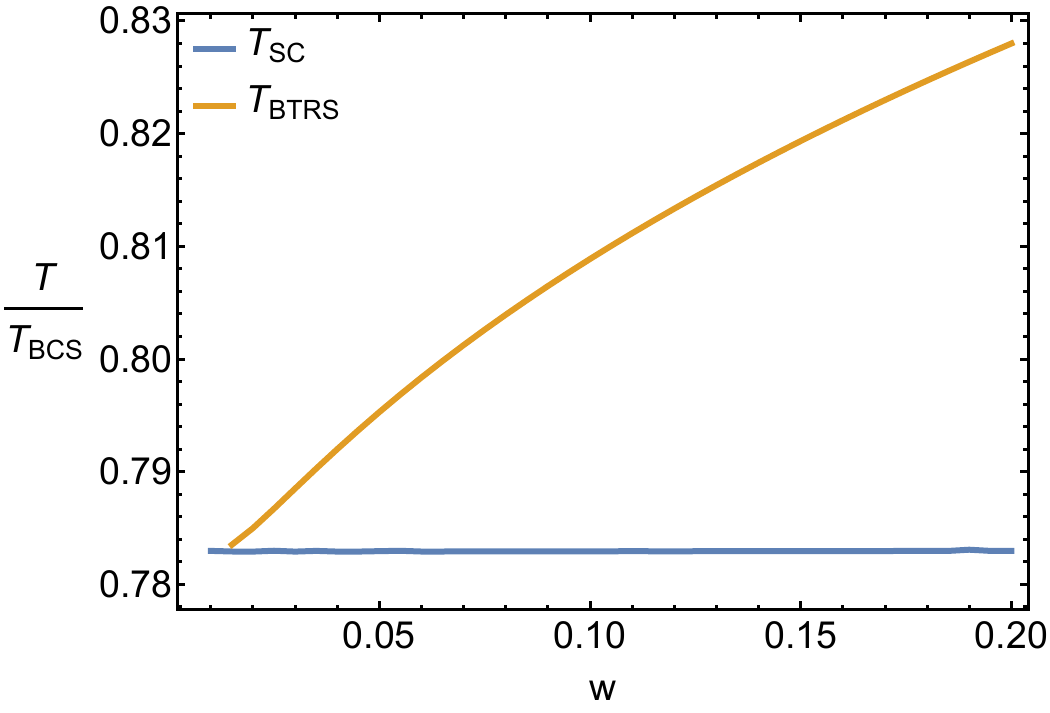}
\caption{
The plot of critical temperatures for transitions into BTRS quadrupling and superconducting states as a function of density-density interaction strength $w$ for $f = h = s = 0.2$.
Note that for sufficiently small $w$ quadrupling phase disappears in this model.
}
\label{fig_tcs_w}
\end{figure}

Next we can expand energy functional \eqref{Omega_L_-d} around the symmetric solution $L_0$, which gives the smallest term proportional to $L$'s difference equal to $\left( \frac{L_+ - L_-}{2} \right)^2 \left( - 4 \Lambda_- + \frac{\kappa}{(2 \sinh(L_0 / 2))^2} \right)$.
Hence, the system transitions from normal to BTRS quadrupling state when $L_0 > L_0^{BTRS}$:
\begin{equation}\label{L0BTRS}
L_0^{BTRS} = \ln{\left( \frac{\kappa + 8 \Lambda_- + \sqrt{\kappa (\kappa + 16 \Lambda_-)}}{8 \Lambda_-} \right)}
\end{equation}
So there is a BTRS quadrupling state between normal and superconducting states if $L_0^{max} > L_0^{BTRS}$ or, in other words, if $\Lambda_- > \Lambda_-^{min}$:
\begin{equation}\label{Lambda-min}
\Lambda_-^{min} = \frac{\kappa}{\left(\left. 4 \sinh{\frac{\Lambda_2}{12 \Lambda_3}} \right|_{t \to t_{SC}} \right)^2}
\end{equation}
It means that in \eqref{Omega_L_-d} the term proportional to $\Lambda_-$ is responsible for the formation of the quadrupling state in this model.
Moreover, $\Lambda_- \propto w^2$ and hence \eqref{Lambda-min} also defines the minimal value of density-density interaction $w_{min}$ that leads to the quadrupling phase in this model.
Since $\frac{\Lambda_2}{12 \Lambda_3} \simeq \frac{0.184}{f h t}$ deep in \eqref{BCS_limit} limit both $\Lambda_-^{min}$ and $w_{min}$ are exponentially small and proportional to $e^{-\frac{0.367}{f h t}}$.
See \figref{fig_tcs_w} for a plot of critical temperatures for different $w$.

\subsection{$3d$ system}

Now we consider a three-dimensional system.
In $3d$, the rescaled skeleton grand canonical potential is given by:
\begin{equation}\label{Omega_L_3d}
\begin{gathered}
\bar{\Omega} \equiv \Omega^s \frac{6 \pi^2 \eta^{3/2}}{T V} = \sum_{a = \pm} \bar{\Omega}_a^0 + \Lambda_2 (L_+ + L_-)^2 \\
- \Lambda_- (L_+ - L_-)^2 - \Lambda_3 (L_+ + L_-)^3 \\
\text{with} \\
\bar{\Omega}_\pm^0 = \kappa^{3/2} \left( \ln\left( \epsilon_\pm + \kappa \right) - 2/3 \right) - \sqrt{\kappa} \left( \epsilon_\pm - 3 \epsilon \right) L_\pm
\end{gathered}
\end{equation}
where $L_a$ again corresponds to the rescaled number of electron pairs in a bosonic component $a$, $\epsilon_\pm = \epsilon + C_\pm / N_F$, and the parameters are defined by:
\begin{equation}
\begin{gathered}
L_\pm = \frac{2 \pi^2 \eta^{3/2} N_F }{\sqrt{\kappa}}\int \frac{d \textbf{q}}{(2 \pi)^3} \widetilde{R}_\pm = 1 - \sqrt{\frac{\epsilon_\pm}{\kappa}} \arctan \sqrt{\frac{\kappa}{\epsilon_\pm}} \\
\eta = \frac{7 \zeta_3 v_F^2}{3 (4 \pi T)^2},\ \ 
\kappa = q_m^2 \eta = \frac{7 \zeta_3 s^2}{3 (4 \pi t)^2} \\
\epsilon = \ln t + \frac{f^2 h^2 s^3 w^2}{12 t} \left( P_2 + P_3 w + P_4 w^2 \right) \\
\Lambda_2 = \frac{(f h s)^2 \sqrt{21 \zeta_3}}{8 \pi} - \sqrt{\frac{3}{7 \zeta_3}} \frac{\pi P_+ f^4 h^4 s^5 w^2}{4 t} \\
\Lambda_3 = \frac{93 \zeta_5 \sqrt{3} f^4 h^4 s^3 t}{112 \pi \zeta_3^{3/2} \sqrt{7}} \\
\Lambda_- = \sqrt{\frac{3}{7 \zeta_3}} \frac{\pi P_- f^4 h^4 s^5 w^2}{4 t}
\end{gathered}
\end{equation}

\begin{figure}
\centering
\includegraphics[width=1\linewidth]{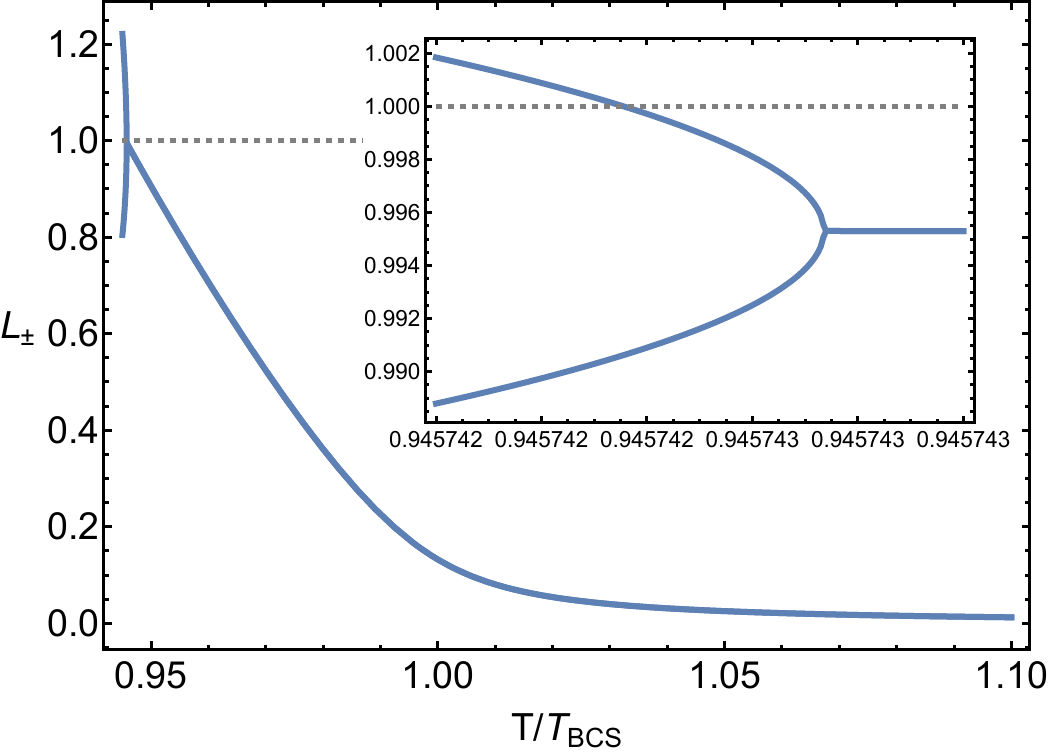}
\caption{
The plot of the rescaled number of electron pairs $L_+,\ L_-$ in the two bosonic components at global minima of the energy function \eqref{Omega_L_-d} as a function of temperature $T$ in a $3d$ system for $f = h = s = 0.5$ and $w = 0.4$.
A single minimum corresponds to a normal state while splitting into two minima to a BTRS quadrupling state.
Note that when $L_\pm = 1$ we get $\epsilon_\pm = 0$ and hence the system transitions to a superconducting state for lower temperatures.
So in this case we have the following phases: superconducting for $t < t_{SC} \simeq 0.945742354$, quadrupling for $t_{SC} < t < t_{BTRS} \simeq 0.945742735$, and normal for $t > t_{BTRS}$.
Cooper pairs are preformed ($L_\pm$ are increased) for $t < t^* \simeq 1$.
}
\label{fig_Ls_t_3d}
\end{figure}

Note, that in this case $L_\pm \in [0, 1]$ for $\epsilon_\pm \in (+\infty,0]$ and hence the $\bar{\Omega}$, \eqref{Omega_L_3d}, doesn't diverge to $-\infty$.
Then normal state corresponds to a single global minimum of $\bar{{\Omega}}$ at $\epsilon_+ = \epsilon_-$ (or equivalently $L_+ = L_-$).
While BTRS quadrupling state appears when there are two global minima with $\epsilon_+ \neq \epsilon_-$.
The system transitions into a superconducting state when energy is minimized for $\epsilon_\pm = 0$ (for one or both components).
See \figref{fig_Ls_t_3d} for a plot of such a transition.
For the $3d$ model considered here, in the relatively weak-coupling limit \eqref{BCS_limit}, the quadrupling state appears only for a very narrow range of temperatures.
Indeed, in this model, the size of this state on the phase diagram should be negligible/vanish in the BCS limit.
Hence, we plot all the quantities somewhat away from that limit.
We have verified that $\epsilon_+ \neq \epsilon_-$ solutions are indeed minima by analytically expanding near the minimum and numerically by the second partial derivative test.

If density-density interaction is sufficiently small, there is no quadrupling state in this model.
In that case, the system transitions into superconducting from the normal state at $\epsilon_+ = \epsilon_- = 0$ and hence:
\begin{equation}
\epsilon = \epsilon_{SC} = - \frac{4 (\Lambda_2 - 3 \Lambda_3)}{3 \sqrt{\kappa}}
\end{equation}
In this case $T_{SC} \to T_{BTRS}$ in \eqref{BCS_limit} limit.
Thus, in this limit quadrupling phase becomes very narrow in temperature since it needs an increased pair propagator.

The developed theory allows calculations of properties that were inaccessible in the previous classical field theory-based approaches.
Below we give two example calculations.

\section{Specific heat}

\begin{figure}
\centering
\includegraphics[width=1\linewidth]{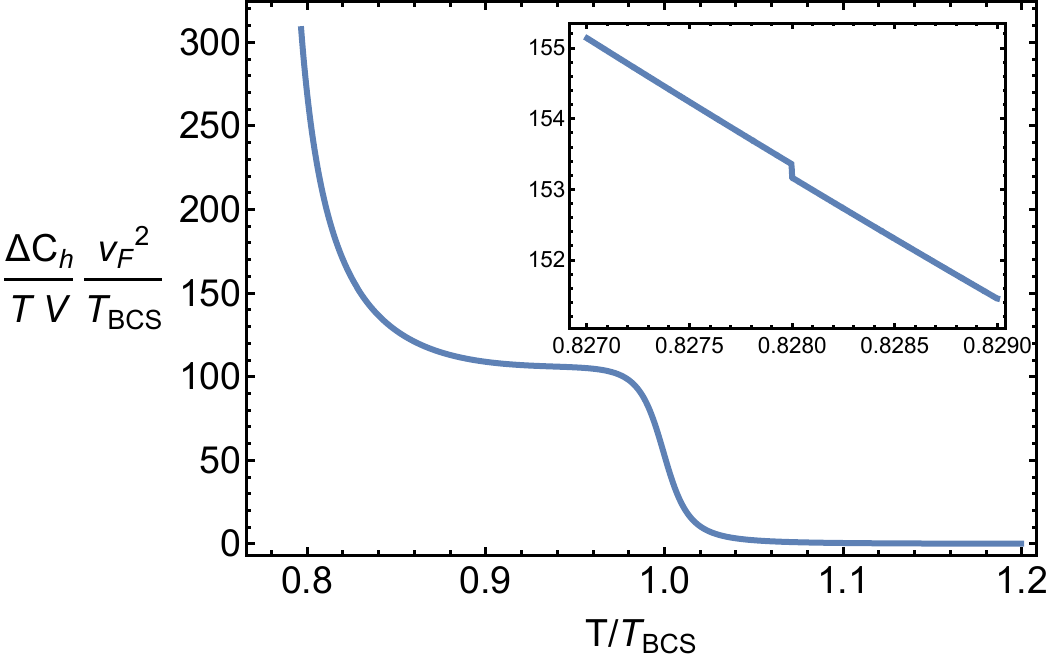}
\includegraphics[width=1\linewidth]{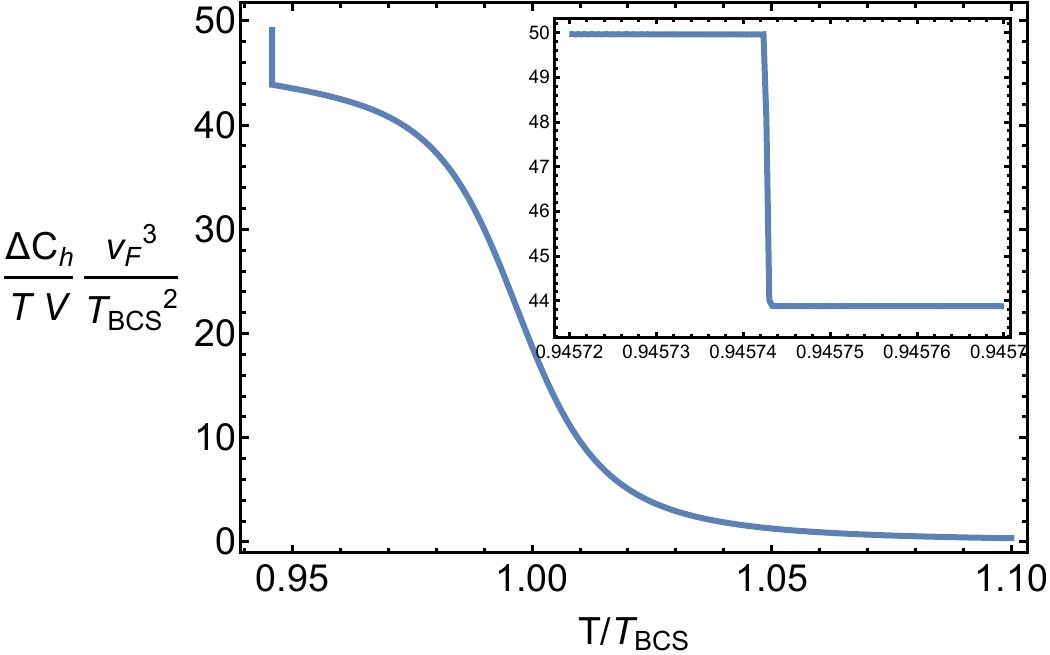}
\caption{
The plot of the rescaled specific heat due to electron pair propagators as a function of temperature for \textbf{(top)} $2d$ model with $f = h = s = 0.2$ and $w = 0.2$, \textbf{(bottom)} $3d$ model with $f = h = s = 0.5$ and $w = 0.4$.
Full-scale plots and zoomed-in part at $T_{BTRS}$ are presented.
As the temperature is lowered, the specific heat is increased due to preformed Cooper pairs at $T^*$.
At the onset of the quadrupling state $T_{BTRS}$, there is a jump.
At least in the weak-coupling models that are considered, the feature at the quadrupling transition is quite small relative to the dominant contribution arising from preformed electron pairs.
}
\label{fig_Ch}
\end{figure}

The previous classical Monte-Carlo calculations assessed contributions of BTRS transition to specific heat, relative to superconducting transition in a London model \cite{shipulin_calorimetric_2023}.
For the interpretation of experiments, it is crucially important to estimate the size of this feature, beyond phase-only models, e.g., relative to the non-singular contribution associated with local pair-formation crossover.
We calculate the specific heat due to the change in electron pair propagators by:
\begin{equation}\label{Ch}
\Delta C_h = - T \frac{\partial^2 \Omega^s}{\partial T^2}
\end{equation}
which results in the following plots for $2d$ and $3d$ systems \figref{fig_Ch}.

Note that in a full critical theory of a $Z_2$ transition, mean-field jumps are converted to a singularity.
Nevertheless, the above calculation suggests that, at least in the moderately weak coupling limit, the specific heat feature at the quadrupling transition is quite small relative to the non-singular background associated with pairing crossover.
This observation is natural, since here quadrupling is associated with establishing coherence of a number of non-condensed electron pairs in different components and hence should be much smaller compared to a specific heat jump arising in BCS models.
Note that experiments on \bkfa \ also reported an extremely small feature at the suggested BTRS transition relative to a nonsingular background associated with pair formation crossover \cite{grinenko_state_2021,shipulin_calorimetric_2023}.

\section{Density of states}

\begin{figure}
\centering
\includegraphics[width=1\linewidth]{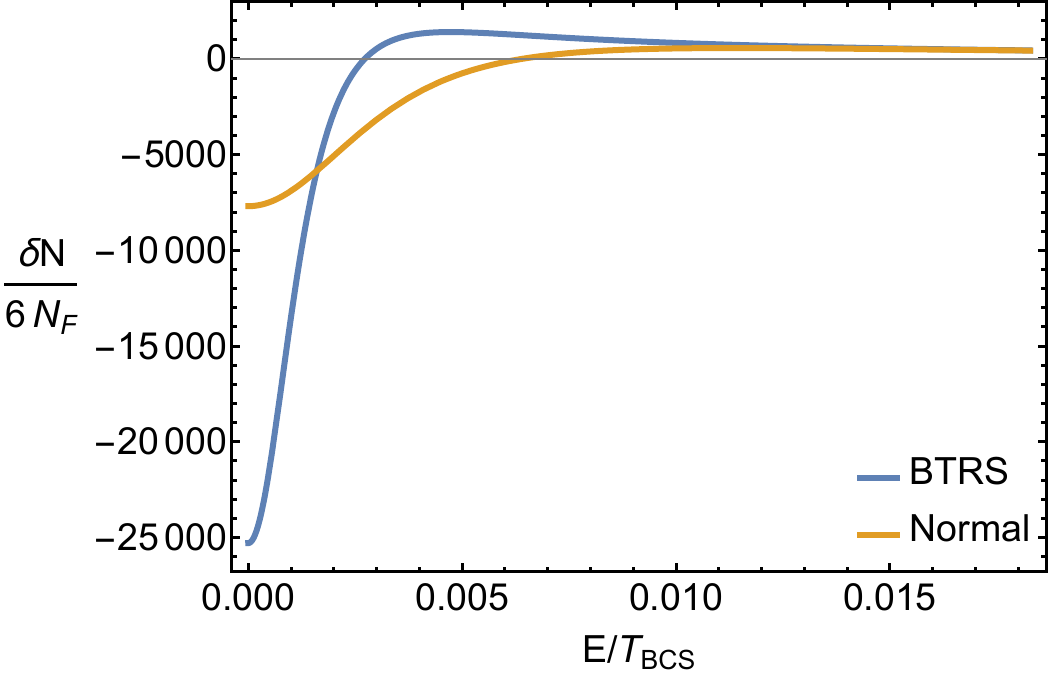}
\includegraphics[width=1\linewidth]{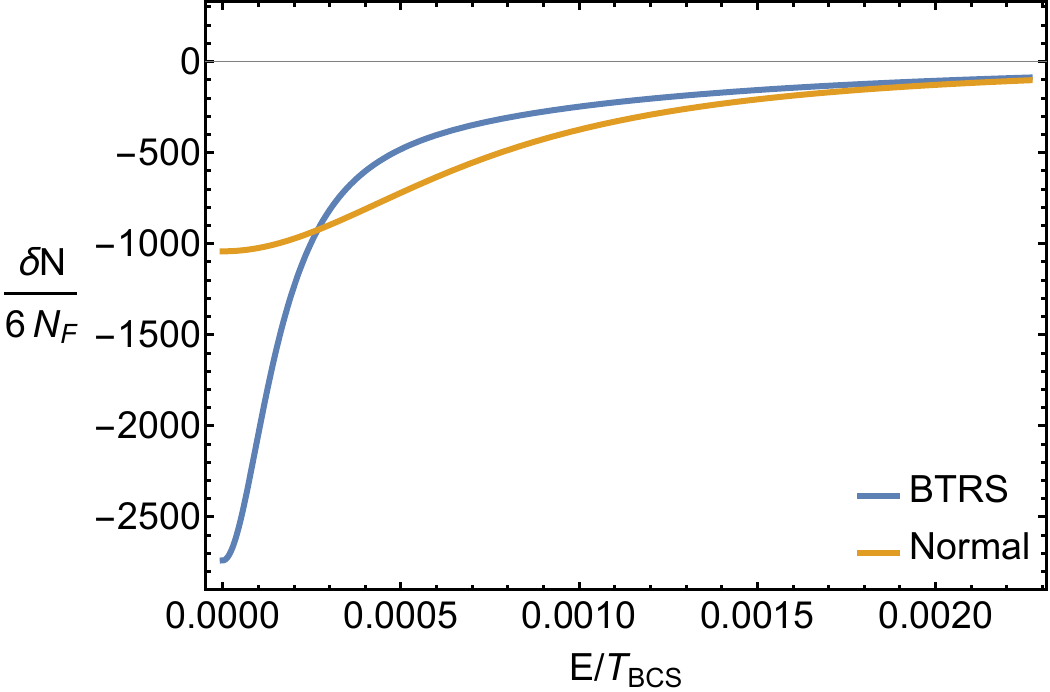}
\caption{
The lowest order estimates for DOS due to pair propagator as function of energy $E$ (it is symmetric for $E < 0$) for \textbf{(top)} $2d$ model with $f = h = s = 0.2$, $w = 0.2$, and $t = 0.82$, \textbf{(bottom)} $3d$ model with $f = h = s = 0.5$, $w = 0.4$ and $t = 0.9457425$.
For these parameters system is in the BTRS quadrupling state with $L_+ \neq L_-$, and the corresponding DOS is marked BTRS.
The plot marked normal is DOS of a reference state for the same parameters but for equal $L$'s given by $(L_+ + L_-) / 2$, which means a state with pre-formed electron pairs but no broken symmetries.
Note that for the quadrupling state, the DOS dip is deeper and narrower.
}
\label{fig_DOS_E}
\end{figure}

Many responses of a system are determined by the density of states (DOS) at the Fermi level.
Let us calculate the change $\delta N$ due to the increase of the pair propagator as a function of energy $E$:
\begin{equation}\label{DOS_def}
\delta N(E) = - \frac{1}{\pi} \text{Im} \sum_{\sigma \alpha} \int \frac{d \textbf{k}}{(2 \pi)^d} \delta\widetilde{G}_{\sigma \alpha}\left( E + \ii 0^+, \textbf{k} \right)
\end{equation}
where $\delta\widetilde{G}_{\sigma  \alpha}\left(\ii \omega_n, \textbf{k}\right)$ is an expansion of a Matsubara frequency and momentum space Green's function to non-zero orders in the pair propagator.
In this work, we will use only the lowest order in $R$, which gives the following estimate for total DOS $\delta N = \delta N_+ + \delta N_-$ for $2d$:
\begin{equation}\label{deltaN_2d}
\begin{gathered}
\frac{\delta N_\pm(E)}{6 N_F} = - \frac{f h t}{3 \epsilon_\pm} D_2\left( \frac{E}{T} \sqrt{\frac{7 \zeta_3}{8 \pi^2 \epsilon_\pm}} \right) \\
D_2(x) =  \left( \frac{1}{1 + x^2} - \frac{x \ln\left[ x+ \sqrt{1 + x^2} \right]}{ (1 + x^2)^{3/2} } \right)
\end{gathered}
\end{equation}
and for $3d$:
\begin{equation}\label{deltaN_3d}
\begin{gathered}
\frac{\delta N_\pm(E)}{6 N_F} = - \frac{4 c^{3/2} (f h t)^2}{3} D_3^{\pm}\left( \frac{E}{T} \frac{\pi}{8} \right) \\
D_3^{\pm}(x) = \text{Re} \frac{1}{\left( \epsilon_\pm - \ii x + \sqrt{\epsilon_\pm (c - 2 \ii x + \epsilon_\pm)} \right) \sqrt{c - 2 \ii x + \epsilon_\pm}} \\
\text{with}\ \ c = \frac{3 \pi^4}{112 \zeta_3}
\end{gathered}
\end{equation}
where we normalize to $6 N_F$ since electrons have $2$ spins and $3$ bands.
These formulas give an estimate of the leading contribution to DOS for small $\epsilon_\pm$ and $E / T_{BCS}$.
Analogous estimates for the single-component case can be found in  \cite{clean_dirty_SC_Fluc_DOS}.
See \figref{fig_DOS_E} for plots of the resulting DOS as a function of energy for $2d$ and $3d$ models.

\begin{figure}
\centering
\includegraphics[width=1\linewidth]{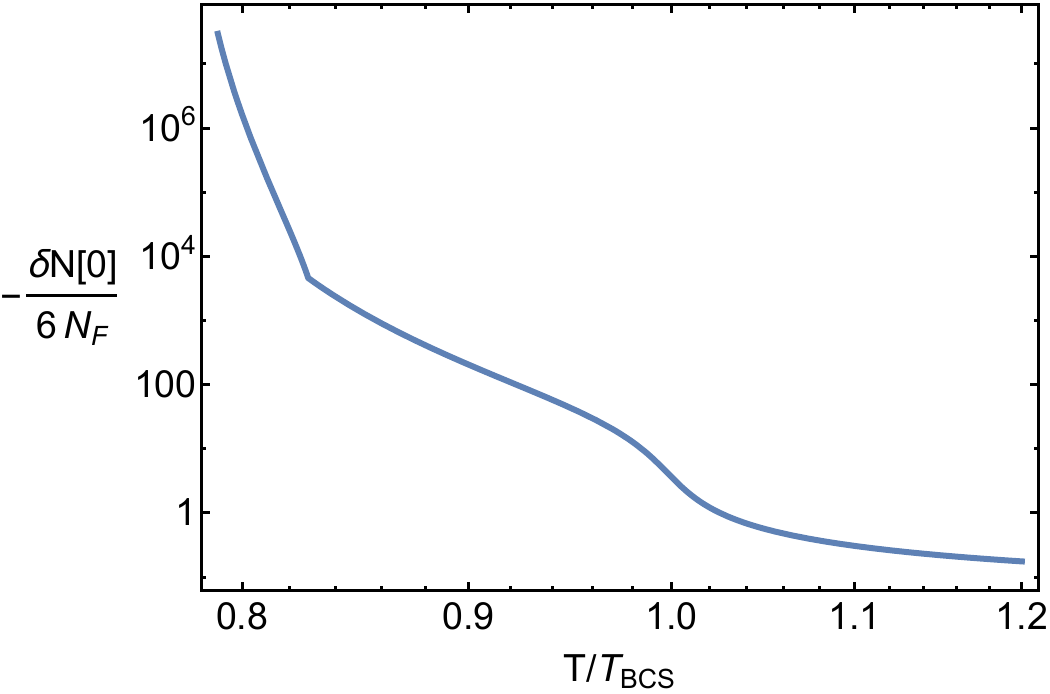}
\includegraphics[width=1\linewidth]{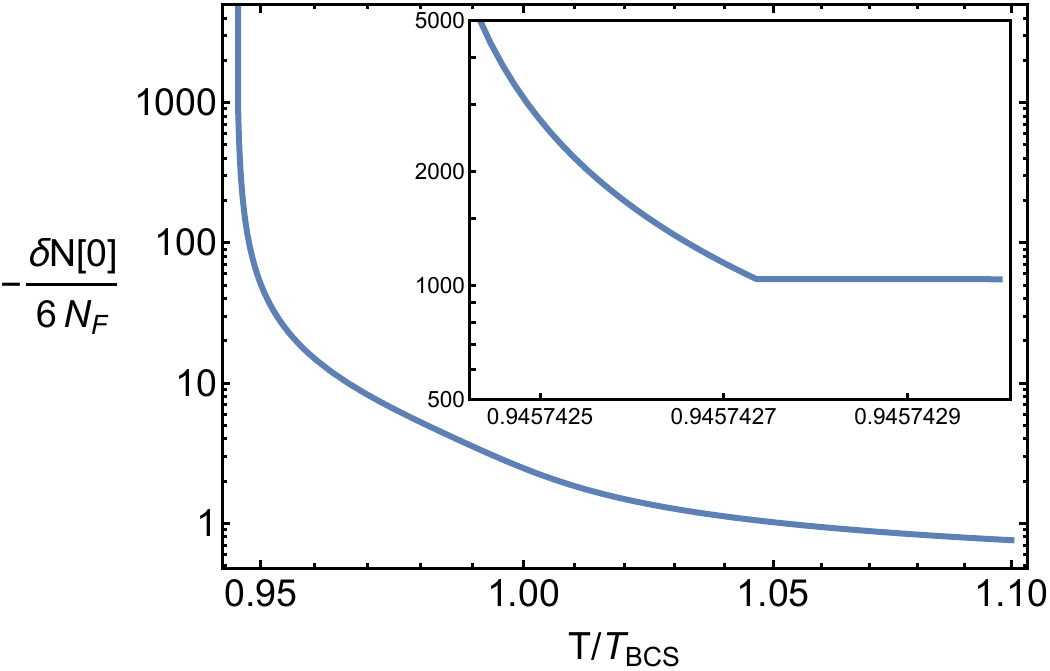}
\caption{
Log scale plot of the lowest order estimates for drop of DOS at Fermi level $\delta N(0)$ as a function of temperature for \textbf{(top)} $2d$ model with $f = h = s = 0.2$ and $w = 0.2$, \textbf{(bottom)} $3d$ model with $f = h = s = 0.5$ and $w = 0.4$.
Note that the DOS drop increases as the temperature is lowered, and the rate increases around $T^* \simeq T_{BCS}$.
At $T_{BTRS}$, there is a kink, which could be a detectable signature of a normal to quadrupling state transition.
}
\label{fig_DOS0}
\end{figure}

We also consider how the change in DOS at the Fermi level depends on temperature.
It is given in $2d$ by:
\begin{equation}
\frac{\delta N(0)}{6 N_F} = - \sum_a \frac{f h t}{3 \epsilon_a}
\end{equation}
and in $3d$ by:
\begin{equation}
\frac{\delta N(0)}{6 N_F} = - \frac{4 c^{3/2} (f h t)^2}{3} \sum_a \frac{1}{\left(  \epsilon_a + \sqrt{\epsilon_a (c + \epsilon_a)} \right) \sqrt{c + \epsilon_a}}
\end{equation}
For the resulting plots, see \figref{fig_DOS0}.
In the normal state, the rate of change with temperature of this DOS is set by $L_+ = L_-$ as a function of temperature.
When the system transitions into the BTRS state, DOS at zero energy will be dominated by the component with a larger $L_a$, and hence its rate of change with temperature will be different.
So this splitting of $L_a$ at $t_{BTRS}$, (see \figref{fig_Ls_t_2d} and \figref{fig_Ls_t_3d}) produces a kink in DOS as function of temperature \figref{fig_DOS0}.

Note that this kink in DOS at the Fermi level should also influence the DOS contribution to conductivity.
Namely, it is expected that this DOS conductivity will be depleted more in the quadrupling state than in the normal ``pseudogap" state.
That feature is potentially resolvable in high-precision angularly resolved photoemission spectroscopy and scanning tunneling spectroscopy.

\section{Conclusions}

In the first part of the paper, we presented a general microscopic fermionic model that allows us to describe the formation of multifermionic orders beyond pairing.
Such orders include various $\langle f_x f_y f_z^\dagger f_w^\dagger \rangle$ condensates as well as charge-4e condensates $\langle f_x f_y f_z f_w \rangle$, which can also be represented in terms of bosonic fields corresponding to electron pairs by $\langle b_x b_y^\dagger \rangle$ and $\langle b_x b_y \rangle$ correspondingly.
Here, the indices $x,y,z,w$  represent generic spin, band, and space indices.

In the second part of the paper, motivated by recent experiments, we specialize to the electron quadrupling condensate that breaks time-reversal symmetry and is characterized by the order parameter composed of linear combination of $\langle f_{\uparrow \alpha} f_{\downarrow \alpha} f_{\downarrow \beta}^\dagger f_{\uparrow \beta}^\dagger \rangle$.
We show that these four-fermionic states arise in microscopic models out of interactions quartic in fermionic operators by considering a concrete microscopic example where the BTRS quadrupling state is stabilized due to an additional repulsive density-density interaction.
Namely, we show, at the level of a microscopic theory, that above the superconducting critical temperature, the symmetry between the preformed but not condensed electron pairs of $s + \ii s$-type and its time-reversed counterpart $s - \ii s$-type can be spontaneously broken: a transition described by a four-electron order parameter.
Overall, we find that tuning the system further away from the weak coupling limit \eqref{BCS_limit} leads to a more pronounced BTRS quadrupling state.
The studied model with $3$ symmetric electronic bands can be easily generalized to higher number of asymmetric electronic bands or bosonic components.

In this state, the system is described by having non-condensed electron pairs and quasiparticles, which are important aspects for understanding unconventional transport and thermal properties of this state.
The formulated model provides a framework for future calculations in generalized mean-field approximations of previously inaccessible physical quantities that depend on electronic spectral functions.
We believe that the behavior of many of the quantities will be qualitatively similar to that in more advanced models that would take into account, for example, intercomponent gauge-field coupling. 
The approach also provides a framework for numerical evaluation of higher-order diagrams using diagrammatic Monte-Carlo as was done for other systems with a broken symmetry state \cite{cdet_antiferr,cdet_BCS}.

\begin{acknowledgments}
We thank Boris Svistunov, Johan Carlstrom, Anton Talkachov, Mats Barkman, Sahal Kaushik, Alexandru Golic, and Igor Timoshuk for useful discussions.
AS and EB were supported by Swedish Research Council Grant 2022-04763, and by Olle Engkvists Stiftelse Grant No. 226-0103.
EB was supported by the Knut and Alice Wallenberg Foundation project KAW 2024.0131 and partially by the Wallenberg Initiative Materials Science for Sustainability (WISE) funded by the Knut and Alice Wallenberg Foundation.
\end{acknowledgments}


%

\clearpage
\appendix

\section{Relating bosonic and fermionic mean-fields corresponding to superconductivity}\label{app_SC_b_f}

Note that, in contrast to BCS theory, here two of the mean-fields $d$ and $\Delta$, or equivalently propagators $\Psi$ and $F$, describe superconductivity.
There is indeed a relation between them, and it is possible to remove the $d$ mean-field.

Using the derivation from the Section II in the manuscript, we see that $\left( \Omega^s - \Omega_0 \right) / T$ will have terms that depend on $\Psi$ and $d$ and their conjugates given by:
\begin{equation}
\begin{gathered}
- C_{x y} \Psi_x^* \Psi_y - P_{x y} \Psi_x^* \Psi_y^* - P_{x y}^* \Psi_x \Psi_y \\
- d_x \Psi_x^* - d_x^* \Psi_x + V_{x y z} \Psi_x^* F_{y z} + V_{x y z}^* \Psi_x F_{y z}^*
\end{gathered}
\end{equation}
Hence, the self-consistency equation for $d$ becomes:
\begin{equation}\label{MFs_eq_d}
d_x = V_{x y z} F_{y z} - C_{x y} \Psi_y - 2 P_{x y} \Psi_y^*
\end{equation}
Next we can use the definition of $\Psi$, which when evaluated gives:
\begin{equation}\label{d_Psi}
d_x = - (E_{x y} + C_{x y}) \Psi_y - 2 P_{x y} \Psi_y^*
\end{equation}
which inserted into \eqref{MFs_eq_d} produces the resulting relation:
\begin{equation}\label{Psi_F}
\Psi_w = - r_{w x} V_{x y z} F_{y z}
\end{equation}
So we obtain that we don't need to solve separately for bosonic superconducting mean-fields $d$ or $\Psi$ -- they are set by fermionic mean-fields through \eqref{Psi_F}.
Moreover, we can insert this solution back into the full skeleton grand canonical potential, which transforms all the terms in $\Omega^s / T$ dependent on $d$, $\Psi$, and their conjugates into:
\begin{equation}\label{d_removed}
\begin{gathered}
- (E_{x y} + 2 C_{x y}) \Psi_x^* \Psi_y - 2 P_{x y} \Psi_x^* \Psi_y^* - 2 P_{x y}^* \Psi_x \Psi_y \\
- d_x \Psi_x^* - d_x^* \Psi_x + V_{x y z} \Psi_x^* F_{y z} + V_{x y z}^* \Psi_x F_{y z}^* = \\
= - v_{x y z w} F_{x y}^* F_{z w}
\end{gathered}
\end{equation}
The resulting term is the same as the initial model without auxiliary bosons would have produced.
So, overall, we can remove the $d$ mean-field completely, but we must add the last term in \eqref{d_removed} to the model.
This shows that this model with $d$ removed (and hence also the model with both $d$ and $\Delta$) will produce the same description of superconductivity as the initial fermionic model would have.
So, for example, setting $W = 0$ and all the mean-fields (except for $d$ and $\Delta$) to zero, and expanding to second order in $\xi$ will produce the usual BCS model.

\section{Offdiagonal pair propagator terms}\label{app_R_offdiag}

Firstly, for brevity we introduce $\epsilon_i$ for $i \in [0,3]$ such that $\epsilon_{a b} = \sum_{i = 0}^3 (\sigma_i)_{a b} \epsilon_i$, where $\sigma_i$ are Pauli matricies.
Namely, $\epsilon_0 = (\epsilon_{1 1} + \epsilon_{2 2}) / 2$, $\epsilon_1 = \text{Re} \epsilon_{1 2}$, $\epsilon_2 = -\text{Im} \epsilon_{1 2}$, and $\epsilon_3 = (\epsilon_{1 1} - \epsilon_{2 2}) / 2$.
Then diagonalizing $\epsilon_{a b}$ (and hence all functions that depend on it) produces its eigenvalues:
\begin{equation}\label{eps_pm}
\epsilon_\pm = \epsilon_0 \pm \sqrt{\epsilon_1^2 + \epsilon_2^2 + \epsilon_3^2}
\end{equation}
It allows us to calculate the rescaled skeleton grand canonical potential:
\begin{equation}\label{Omega_offdiag}
\begin{gathered}
\bar{\Omega} = \sum_{a = \pm} \bar{\Omega}_a^0 + \Lambda_2 \left( L_+ + L_- \right)^2 \\
+ \left[ - 4 \Lambda_- \epsilon_3^2 + 2 \left( \Lambda_2 + \Lambda_- \right)  (\epsilon_1^2 + \epsilon_2^2) \right] \left( \frac{L_+ - L_-}{\epsilon_+ - \epsilon_-} \right)^2 \\
- \Lambda_3 \left[ (L_+ + L_-)^3 + 2 \epsilon_1 (\epsilon_1^2 - 3 \epsilon_2^2) \left( \frac{L_+ - L_-}{\epsilon_+ - \epsilon_-} \right)^3 \right. \\
\left. + 6 (\epsilon_1^2 + \epsilon_2^2) (L_+ + L_-) \left( \frac{L_+ - L_-}{\epsilon_+ - \epsilon_-} \right)^2 \right]
\end{gathered}
\end{equation}
where $\bar{\Omega}_\pm^0$ and $L_\pm$ depend on $\epsilon_\pm$ correspondingly, as described in the main text.

Note, that all terms in \eqref{Omega_offdiag}, exept one depend only on $\epsilon_1^2 + \epsilon_2^2$.
This term is proportional to $\epsilon_1 (\epsilon_1^2 - 3 \epsilon_2^2)$ with some non-negative prefactor (since for $\epsilon_+ > \epsilon_-$ we have $L_+ < L_-$).
So in minimum we will have either $\epsilon_1 = \epsilon_2 = 0$ or they will break some $Z_3$ symmetry: $\epsilon_1 + \ii \epsilon_2 = e^{\ii \frac{\pi}{3} (1 + 2 n)}$ for $n = 0, 1, 2$.
So to check whether the last case is possible, we set them to one of the $3$ solutions: $\epsilon_1 < 0$ and $\epsilon_2 = 0$.
Then we expand $\bar{\Omega}$ \eqref{Omega_offdiag} in powers of $\epsilon_1$ around $0$.
When $\epsilon_3 = 0$ is a minimum (the system is in the normal state), we obtain the first non-trivial term in $\epsilon_1$ term:
\begin{equation}
\begin{gathered}
\left. \left[ \frac{\partial_{\epsilon_3}^2 \bar{\Omega}}{2} + 2 \left( \Lambda_2 + 3 \Lambda_- - 6 \Lambda_3 L_0 \right) \left( \partial_{\epsilon_0} L_0 \right)^2 \right] \right|_{\epsilon_1 = \epsilon_3 = 0} \epsilon_1^2
\end{gathered}
\end{equation}
where prefactor is positive since $\partial_{\epsilon_3}^2 \bar{\Omega} > 0$ when $\epsilon_3 = 0$ is a minimum and maximal values of $L_0$ are in $2d$: $L_0^{max} = \frac{\Lambda_2}{6 \Lambda_3}$, in $3d$: $L_0^{max} = 1$.

When $\epsilon_3 \neq 0$ is a minimum (the system is in the BTRS quadrupling state), the first non-trivial in $\epsilon_1$ term is (using that $\partial_{\epsilon_3} \bar{\Omega} = 0$):
\begin{equation}
\begin{gathered}
\left. 2 \left[ \Lambda_2 + 3 \Lambda_- - 3 \Lambda_3 (L_+ + L_-) \right] \left( \frac{L_+ - L_-}{\epsilon_+ - \epsilon_-} \right)^2 \right|_{\epsilon_1 = \epsilon_3 = 0} \epsilon_1^2
\end{gathered}
\end{equation}
By the same logic, we see that the prefactor is positive.
So overall we conclude that the skeleton grand canonical potential is minimized by $\epsilon_1 = \epsilon_2 = 0$ and hence diagonal $\epsilon_{a b}$ (and diagonal pair propagator $R_{a b}$).

\section{Numerical coefficients $P$}\label{app_Ps}

$P$'s depend on $x = \frac{\omega_f}{T} = \frac{1}{h t}$ cutoff and are defined from the summation of the corresponding diagrams.
To that end, let us introduce:
\begin{equation}\label{Gamma_one}
\begin{gathered}
\Gamma_{a b c d k}(x) = \sum_{n = -\infty}^{+ \infty} \int_{-x}^{x} dy\gamma(n, y, a) \gamma(n, y, b)^* \\
\gamma(n + k, y, c) \gamma(n + k, y, d)^*\\
 \\
\text{with} \\
\gamma(k, y, a) = \frac{1}{ ( 2 \pi \ii (k + 1/2) - y)^a}
\end{gathered}
\end{equation}
We will use sums of $\Gamma_{\textbf{v} k}$ where $\textbf{v} = (a, b, c, d)$ defined by:
\begin{equation}\label{Gamma_many}
\Gamma_{\textbf{v} \textbf{u} ...} = \sum_{k = -\infty}^{+ \infty} \Gamma_{\textbf{v} k} \Gamma_{\textbf{u} k} ...
\end{equation}
Then $P$'s are given by:
\begin{equation}
\begin{gathered}
P_2 = 16 ( 4 \Gamma_{0,1,0,1,0,1,1,2} + \Gamma_{0,1,0,1,1,1,1,1} \\
+ 2 \Gamma_{0,1,1,1,0,1,1,1} + 2 \Gamma_{0,1,2,1,0,1,1,0} ) \\
P_3 = - 64 ( \Gamma_{0,1,1,1,0,1,1,1,0,1,0,1} - \Gamma_{0,1,1,2,0,1,0,1,0,1,0,1} \\
- \Gamma_{1,1,1,1,0,1,0,1,0,1,0,1} + \Gamma_{0,1,1,1,0,1,1,1,0,1,1,0} \\
+ \Gamma_{0,1,2,1,0,1,1,0,0,1,1,0}) \\
P_4 = 128 (\Gamma_{0,1,1,1,0,1,0,1,0,1,1,1,0,1,1,0} + \Gamma_{0,1,1,1,0,1,1,1,0,1,0,1,0,1,0,1} \\
+ 13 \Gamma_{0,1,1,2,0,1,0,1,0,1,0,1,0,1,0,1} + \Gamma_{0,1,2,1,0,1,1,0,0,1,1,0,0,1,1,0} \\
+ 6 \Gamma_{1,1,1,1,0,1,0,1,0,1,0,1,0,1,0,1} + \Gamma_{0,1,1,1,0,1,1,1,0,1,1,0,0,1,1,0})
\end{gathered}
\end{equation}

\begin{equation}
\begin{gathered}
P_+ = \frac{2}{3} ( 64 \Gamma_{0,1,0,1,0,1,2,3} + 32 \Gamma_{0,1,0,1,1,1,2,2} \\
+ 26 \Gamma_{0,1,0,1,1,2,1,2} + 32 \Gamma_{0,1,1,1,0,1,2,2,2} \\
+ 46 \Gamma_{0,1,1,2,0,1,1,2} + 20 \Gamma_{0,1,1,2,1,1,1,1} \\
+ 26 \Gamma_{0,1,2,1,0,1,2,1} + 32 \Gamma_{0,1,2,2,0,1,1,1} \\
+ 15 \Gamma_{1,1,1,1,1,1,1,1} - 40 \Gamma_{1,1,1,2,0,1,1,1} \\
+ 32 \Gamma_{0,1,3,2,0,1,1,0} + 10 \Gamma_{1,2,2,1,0,1,1,0}) \\
P_- = 2 ( 2 \Gamma_{0,1,0,1,1,2,1,2}, + 6 \Gamma_{0,1,1,2,0,1,1,2} \\
+ 4 \Gamma_{0,1,1,2,1,1,1,1} + 2 \Gamma_{0,1,2,1,0,1,2,1} + 3 \Gamma_{1,1,1,1,1,1,1,1} \\
- 8 \Gamma_{1,1,1,2,0,1,1,1} + 2 \Gamma_{1,2,2,1,0,1,1,0})
\end{gathered}
\end{equation}
where integrals over $y$ in \eqref{Gamma_one} and sums over $k$ in \eqref{Gamma_many} were evaluated numerically.

\end{document}